\documentclass[11pt]{article}
\usepackage{cite}
\usepackage{amsmath,amsfonts,amssymb}%,calrsfs}
\usepackage[small,bf,hang]{caption}
\usepackage{slashed}
\usepackage{epsfig}

\makeatletter\let\corresponds\@undefined\makeatother

\usepackage[vcentermath]{youngtab}

\def\hybrid{
        \topmargin -20pt
        \oddsidemargin 0pt
        \headheight 0pt \headsep 0pt
        \textwidth 6.25in % A4 paper
        \textheight 9.5in % A4 paper
        \marginparwidth .875in
        \parskip 5pt plus 1pt \jot = 1.5ex}

\hybrid

 \csname
@addtoreset\endcsname{equation}{section}

\def\moth{\mathsurround=0pt}
\newdimen\zo \zo=0pt

\def\tick{\leaders\hrule height 0.5ex depth 0pt \hskip 0.5pt}
\def\upboxfill{$\moth \setbox\zo\hbox{\tick}%
  \hskip 3pt\hbox to 0pt{$\tick$\hss}\hrulefill \hbox to 7.5pt{$\tick$\hss}$}

\def\dtick{\leaders\hrule height .34pt depth 0.5ex \hskip 0.5pt}
\def\downboxfill{$\moth \setbox\zo\hbox{\dtick}%
  \hskip 2pt\hbox to 0pt{$\dtick$\hss}\hrulefill \hbox to 2pt{$\dtick$\hss}$}

\def\bec{\begin{center}}
\def\ec{\end{center}}

\def\p{\partial}

 \def\det{{\rm det\,}}
\def\be{\begin{equation}}
\def\ee{\end{equation}}
\def\bea{\begin{eqnarray}}
\def\eea{\end{eqnarray}}
\def\ba{\begin{array}}
\def\ea{\end{array}}

\def\ket#1{|#1\rangle}

\def\fancys{\mathbb{S}}

\begin{document}

\begin{titlepage}
\rightline{}
\rightline{\tt MIT-CTP-4494}
\rightline{\tt LMU-ASC 59/13}
\rightline{\tt MPP-2013-241}
%\rightline\today
\rightline{September 2013}
\begin{center}
\vskip 1cm
{\Large \bf {The Spacetime of Double Field Theory: \\[0.4cm] Review, Remarks, and Outlook}}\\
\vskip 1.2cm
{\large {Olaf Hohm${}^1$, Dieter L\"ust${}^{1,2}$ and Barton Zwiebach${}^3$}}
\vskip 1cm
{\it {${}^1$Arnold Sommerfeld Center for Theoretical Physics}}\\
{\it {Theresienstrasse 37, D-80333 Munich, Germany}}\\
olaf.hohm@physik.uni-muenchen.de, dieter.luest@lmu.de
\vskip 0.2cm
{\it {${}^2$Max-Planck-Institut f\"ur Physik}}\\
{\it {F\"ohringer Ring 6, 80805 M\"unchen, Germany}}\\
\vskip 0.2cm
{\it {${}^3$Center for Theoretical Physics}}\\
{\it {Massachusetts Institute of Technology}}\\
{\it {Cambridge, MA 02139, USA}}\\
zwiebach@mit.edu

\vskip 0.9cm
{\bf Abstract}
\end{center}

%\vskip 0.01cm

\noindent
\begin{narrower}

We review double field theory (DFT) with emphasis on the doubled  
spacetime and its generalized coordinate  
transformations, which unify  diffeomorphisms   
and $b$-field gauge transformations. 
We illustrate how the composition of generalized 
coordinate transformations fails to associate.  
 Moreover, in dimensional reduction, the $O(d,d)$ T-duality transformations of fields
 can be obtained as generalized diffeomorphisms.
Restricted  to a half-dimensional subspace, DFT includes  
`generalized geometry',  but is more   
general in that local patches of the doubled space may be 
glued together with generalized coordinate transformations. 
Indeed, we show that for certain T-fold  
 backgrounds with non-geometric
fluxes, there are generalized coordinate transformations that induce, 
as gauge symmetries of DFT,
the requisite $O(d,d;\mathbb{Z})$ monodromy transformations.  
Finally we review recent results on the $\alpha'$ extension of DFT which, 
reduced to the half-dimensional subspace, yields intriguing modifications of the
basic structures of generalized geometry.

\end{narrower}

\end{titlepage}

\newpage

\tableofcontents

\newpage

\section{Introduction}
The underlying symmetry of gravity is given by the
diffeomorphism group or the group of 
general coordinate transformations. Accordingly, the basic 
dynamical variables are tensor fields, including the metric tensor $g_{ij}$ that 
describes the spacetime geometry. This holds true in 
the target space description of  
 string theory, where the metric is augmented by 
an antisymmetric tensor $b_{ij}$ (the $b$-field) and a scalar $\phi$ 
(dilaton), as well as  
various $p$-forms, depending on the string theory 
considered. The universal spacetime  low-energy  action for the massless fields  
common 
to all oriented closed 
string theories 
then reads 
 \bea\label{original}
  S \ = \ \int d^Dx \sqrt{-g}e^{-2\phi}\left[R+4(\partial\phi)^2-\frac{1}{12}H^2\right]\,,
 \eea
where $D=26$ or $D=10$, and $H$ denotes the three-form field strength of the $b$-field. 
This action admits the diffeomorphism invariance of gravity and  
an abelian gauge symmetry of the $b$-field, but it does not display  
any `higher' symmetry 
that would explain the special 
role of the $b$-field and dilaton, as opposed to any other matter fields 
that can be coupled to gravity. 
In closed string theory, however, the field content is uniquely determined
and closely related to the T-duality group $O(d,d; \mathbb{Z} )$ 
 that is well known to emerge~\cite{Giveon:1994fu}
 when the theory 
is put on a torus background $T^d$.  
In fact, for the low-energy theory of the massless fields the symmetry becomes
a continuous $O(d,d;\mathbb{R})$, henceforth written as just $O(d,d)$. 
The components of $g_{ij}$
and $b_{ij}$ along the torus transform into each other according to the Buscher rules~\cite{buscher}, 
and it is only the particular action (\ref{original}) that is compatible with T-duality. 
This naturally leads one to wonder if
there is a way to make these features  
manifest at the level of a spacetime action such as  (\ref{original}). In this paper we will 
review `double field theory' that provides such a formulation  
\cite{Siegel:1993th,Hull:2009mi,Hull:2009zb,Hohm:2010jy,Hohm:2010pp}
and has been the focus of much recent attention.  
For earlier attempts see \cite{Tseytlin:1990va,Duff:1989tf,Siegel:19932,Maharana:1992my}, and 
the concluding section for a more detailed guide to the literature.

In double field theory (DFT) 
an action as (\ref{original}) can be 
formulated in an $O(D,D)$ covariant fashion by organizing $g$, $b$ and $\phi$ 
into new field variables that are $O(D,D)$ tensors \cite{Hohm:2010pp}. Specifically, the fundamental 
fields are given by a symmetric $O(D,D)$ matrix 
 \be\label{firstH}
  {\cal H}_{MN} \ = \  \begin{pmatrix}    g^{ij} & -g^{ik}b_{kj}\\[0.5ex]
  b_{ik}g^{kj} & g_{ij}-b_{ik}g^{kl}b_{lj}\end{pmatrix}\;,
 \ee
and an $O(D,D)$ singlet dilaton $d$ related to $\phi$ via $e^{-2d}=\sqrt{-g}e^{-2\phi}$, 
where $M,N=1,\ldots, 2D$ are fundamental $O(D,D)$ indices. The matrix (\ref{firstH})
is naturally viewed as a metric on a doubled space with $2D$ coordinates $X^{M}=(\tilde{x}_i,x^i)$, 
with corresponding derivatives 
$\partial_M = (\tilde\partial^i , \partial_i)$,    
where $x^i$ are the usual spacetime coordinates. 
The additional coordinates $\tilde{x}_i$, dual to winding modes of the closed string,   
 are known to be present in the full string theory on a torus, as seen when formulated 
as a second-quantized string field theory~\cite{Kugo:1992md,Zwiebach:1992ie }.  
In DFT all coordinates are doubled  
for any background, 
while imposing a constraint that effectively renders half of them `inactive'. More precisely, 
we impose the constraint 
 \be\label{STRONG}
  \eta^{MN}\partial_{M}\partial_{N} \ \equiv \ \partial^{M}\partial_{M} \ = \ 0\;, \qquad 
   \eta^{MN} \ = \ \begin{pmatrix}    0 & {\bf 1} \\[0.5ex]
  {\bf 1} & 0 \end{pmatrix}\;, 
  \ee
where   
$\eta_{MN}$ denotes the $O(D,D)$ invariant metric. 
The vanishing of $\partial^{M}\partial_{M} $ when acting on arbitrary fields and parameters is the weak form of the constraint and can be identified with the level
matching constraint of closed string theory.  When, in addition, $\partial^{M}\partial_{M} $ is  constrained to vanish for all products of fields and gauge
parameters, we have the strong version of the constraint.
The strong version goes beyond the  
 level-matching constraint of closed string theory
 and implies that the fields depend only 
on half of the (doubled) coordinates.   
When DFT is applied to a background with
$d$ abelian isometries, like a torus $T^d$, 
and fields are independent of these $d$ coordinates, we can, however, 
realize
the full $O(d,d)$ symmetry geometrically by the use of 
$\tilde{x}$-dependent coordinate transformations.

The usual gauge transformations of the metric and  $b$-field, i.e., diffeomorphisms generated by  a 
vector $\xi^i$ and 
abelian gauge transformations $\delta_{\tilde{\xi}}b_{ij}=\partial_i\tilde{\xi}_j-\partial_j\tilde{\xi}_i$ can 
be lifted to $O(D,D)$ covariant gauge transformations of ${\cal H}_{MN}$ and the dilaton $d$, 
with parameter $\xi^M=(\tilde{\xi}_i,\xi^i)$, 
 \be\label{gendiff}
  \begin{split}
   \delta_{\xi}{\cal H}_{MN} \ &= \
   \xi^{P}\partial_{P}{\cal H}_{MN}+\big(\partial_{M}\xi^{P}-\partial^{P}\xi_{M}\big){\cal H}_{PN}
   +\big(\partial_{N}\xi^{P}-\partial^{P}\xi_{N}\big){\cal H}_{MP}\;, \\
   \delta_{\xi}\big(e^{-2d}\big) \ &= \ \partial_{M}\big(\xi^{M}e^{-2d}\big)\;,
  \end{split}
 \ee
where indices are raised and lowered with $\eta_{MN}$ defined in (\ref{STRONG}). When specializing
to the components of ${\cal H}_{MN}$ in (\ref{firstH})
 and setting $\tilde{\partial}^i=0$ these transformations reduce to the standard gauge 
transformations. Let us stress that also the gauge transformations for $\tilde{\partial}^i \neq 0$ are well 
motivated from string theory. In fact, the gauge transformations to cubic order have been derived from 
closed string field theory~\cite{Hull:2009mi}, 
and there is a unique way to make them background independent as 
transformations of $g_{ij}$ and $b_{ij}$~\cite{Hohm:2010jy}.   
Application to  (\ref{firstH}) then uniquely leads to the above gauge 
transformations of ${\cal H}$ and $d$.  

There is a unique scalar ${\cal R}$ 
written in terms of second derivatives of ${\cal H}_{MN}$ and $d$ 
that indeed transforms as a scalar under  (\ref{gendiff}), i.e., $\delta_{\xi}{\cal R}=\xi^M\partial_M{\cal R}$.
Since $e^{-2d}$ transforms as a density, the scalar ${\cal R}$ can be used
to write a manifestly gauge invariant action 
 \be\label{gaugeinvaction}
  S_{\rm DFT} \ = \ \int d^{2D}X\,e^{-2d}\,{\cal R}({\cal H},d)\;.
 \ee
Setting $\tilde{\partial}^i=0$ and writing it in terms of $g$, $b$ and $\phi$ the
above action reduces to (\ref{original}). 
Thus, once formulated in terms of the right dynamical objects and geometrical structures, the 
two-derivative part of the spacetime action of bosonic string theory is unique, 
quite in contrast to the original formulation. Again, for $\tilde{\partial}^i \neq 0$,
 this action is also 
well motivated from string theory in that  
expanded to cubic order around 
a constant toroidal  
background, it  coincides with the closed string field theory action of the massless fields to that order. 

In this article 
we will elaborate on the geometrical implications of DFT.  
It is mainly a review, but we also give some 
new results. 
 Specifically, we show in sec.~4 that even the strongly constrained DFT allows for backgrounds that are globally well-defined in the doubled geometry, but not in standard differential geometry,  
in this sense going beyond conventional supergravity. 
We will discuss the role of generalized coordinate transformations on 
the doubled space and why they require some notion of generalized manifold. 
To explain this point recall that in general relativity 
we begin with a conventional manifold for which coordinates on overlapping patches are related 
by the usual general coordinate transformations. These, in turn, can be written infinitesimally as
Lie derivatives generated by the vector parameter $\xi^i$. In contrast, the gauge transformations 
(\ref{gendiff}) are not given by Lie derivatives on the doubled space but rather represent `generalized 
Lie derivatives' $\widehat{\cal L}_{\xi}$, 
so that we have $\delta_{\xi}{\cal H}_{MN}=\widehat{\cal L}_{\xi}{\cal H}_{MN}$. If follows immediately that 
we cannot view (\ref{gendiff}) as infinitesimal diffeomorphisms on the doubled space and, therefore, 
the doubled space needs to be viewed as a suitably generalized manifold that is `patched together'
by some generalized coordinate transformations.  

A closely related observation is that the formulation 
of DFT requires the \textit{constant} $O(D,D)$ metric $\eta_{MN}$ defined in (\ref{STRONG}). 
For a conventional manifold there is 
no coordinate independent sense in which a metric 
can take the constant form in (\ref{STRONG}). Put differently, the doubled manifold would be of a rather special 
`flat' form, allowing 
for preferred coordinate systems for which $\eta_{MN}$ is constant. 
In contrast, the generalized Lie derivative in DFT does leave $\eta_{MN}$ invariant, $\widehat{\cal L}_{\xi}\eta_{MN}=0$, 
implying that the notion of coordinate transformations on the doubled space should be generalized in 
such a way that $\eta_{MN}$ is actually invariant. 
 
In a recent paper, two of us gave a proposal for such generalized coordinate transformations that meet all 
consistency conditions tested so far. A generalized vector $V_M$ transforms 
under $X\rightarrow X'$ as~\cite{Hohm:2012gk}: 
 \be\label{GenTRAns}
  V'_M(X') \ = \ {\cal F}_M{}^{N} V_{N}(X)\;,
 \ee 
where  
 \be
 \label{GenTRAF}
  {\cal F}_{M}{}^{N} \ = \ \frac{1}{2} \left(   \frac{\partial X^{P}}{\partial X'^{M}}\,\frac{\partial X^{\prime}_{P}}{\partial X_{N}}
  +\frac{\partial X^{\prime}_{M}}{\partial X_{P}}\, \frac{\partial X^{N}}{\partial X'^{P}}\right)\;, 
 \ee 
and indices on coordinates are raised and lowered with $\eta_{MN}$.  
Similarly, an arbitrary generalized tensor transforms tensorially, each index being rotated by ${\cal F}$. 
Even though these transformations do not describe conventional general coordinate 
transformations on the $2D$-dimensional space they 
{\em do} encode arbitrary general coordinate transformations 
on $D$-dimensional isotropic subspaces of the doubled space, that is, spaces for which tangent vectors
are null in the metric $\eta$. 
The components of ${\cal H}_{MN}$ in (\ref{firstH}) then transform 
conventionally as tensors, 
without imposing any constraints on the geometry encoded by the $D$-dimensional 
metric $g_{ij}$. 
As we will review, at the same time the action of ${\cal F}$ 
on $\eta_{MN}$ is such that it is left invariant, 
as required. For fields depending only on $x$, 
 $b$-field gauge transformations 
are generalized coordinate transformations $\tilde{x}_i\rightarrow \tilde{x}_i-\tilde{\xi}_i(x)$
that mix $x$ and $\tilde{x}$ coordinates.

DFT is a framework 
that is flexible enough to encode all that is contained in the usual spacetime action (\ref{original}). 
In particular,  the presence of the `flat' $O(D,D)$ metric $\eta_{MN}$ does not imply that the spacetime 
metric $g_{ij}$ (encoded by ${\cal H}_{MN}$) is also flat or, for that matter, restricted at all. In fact, 
DFT is also flexible enough to encode the 
framework of   
generalized geometry, which has been developed 
in pure mathematics~\cite{Hitchin:2004ut,Gualtieri:2003dx,Gualtieri:2007bq,Vaisman:2012ke}. 
This geometry does not change the underlying manifold $M$ (it is not doubled). 
The tangent bundle $T(M)$, however, is replaced  by $T(M)\oplus T^*(M)$ and  structures such as the 
Courant bracket are defined on this extended bundle. 
Before the advent of DFT, however,  it appears that generalized geometry was not  
 developed to the extent that invariant curvatures and thus actions such as (\ref{gaugeinvaction}) could be defined. 
  Generalized geometry is   
  manifestly  contained in DFT in that we may solve the strong constraint (\ref{STRONG})
by setting, say, $\tilde{\partial}^i=0$, after which the components $(V_i,V^i)$ of a generalized vector 
$V^M$  acquire a definite interpretation as vector ($V^i$) and one-form ($V_i$),   
thereby encoding an element of $T\oplus T^*$. Moreover, 
the generalized Lie derivatives
of vectors   
are given by the action of the so-called Dorfman bracket, and their closure is 
governed by the Courant bracket.  
In this way, to zeroth order in $\alpha'$, DFT may be viewed 
as the first implementation of  generalized geometry at the 
level of the full spacetime action. However, as we will also review, 
taking $\alpha'$ to be non-zero the generalized Lie derivative of DFT acquires  $\alpha'$ 
corrections that even on the half-dimensional subspace modify the defining structures of 
generalized geometry~\cite{Hohm:2013jaa}.

Apart from implementing the generalized geometry program and thus 
reformulating the usual spacetime theory in terms of a geometry that is better adapted 
to the T-duality properties of string theory, we will   
show how even at the two-derivative level DFT is yet more general, 
naturally encoding `non-geometric' backgrounds. The most direct way to describe backgrounds 
that are not captured by ordinary supergravity is to try to relax the strong constraint (\ref{STRONG})
so that solutions may depend locally both on $x$ and $\tilde{x}$. In fact, in the full closed string field 
theory on a torus background only the weaker level-matching constraint is required. 
It is a subtle question whether the constraint can be relaxed for the massless fields only (or to zeroth order in $\alpha'$) or for different backgrounds, but there 
has been progress   
exhibiting relaxed constraints  
in massive deformations of type IIA~\cite{Hohm:2011cp}, generalized 
Scherk-Schwarz compactifications \cite{Aldazabal:2011nj,Geissbuhler:2011mx}, 
and now even  
more generally in \cite{Geissbuhler:2013uka}. 
We will be 
assuming the strong constraint throughout 
the paper.
We will 
see that even in the strongly constrained DFT 
there are still `non-geometric' field configurations in DFT that are globally well-defined   
when patched with the generalized coordinate transformations (\ref{GenTRAns}). These examples 
are closely related to the idea 
of `T-folds'~\cite{Hull:2004in},  
where one allows `patching by $O(d,d)$ transformations'.
We believe, however, that a more precise formulation of this idea, in the sense of a natural extension 
of the general coordinate transformations of differential geometry, was lacking. 
We will argue that (\ref{GenTRAns}) fills the gap. 

For the convenience of the reader we summarize in the following the main messages of 
the review part:
 \begin{itemize}
 \item 
  DFT can be viewed, in particular, as the physical implementation of the 
   `generalized geometry' concepts of Hitchin-Gualtieri. In particular, it encodes the familiar 
 low-energy limits of string theory 
 in generality: the backgrounds are not restricted to tori.
 DFT and `generalized geometry' are not really `alternative' approaches. 
 \item The gauge transformations of the low-energy limit
 of string theory are unified in DFT in the form of \textit{generalized 
 coordinate transformations}.  These treat (arbitrary) diffeomorphisms of the 
 $D$-dimensional subspace and $b$-field gauge transformations on the same footing, in addition 
 to encoding $O(d,d)$ transformations for certain configurations. Generalized coordinate transformations  
 are quite different from ordinary diffeomorphisms:  they compose in a non-standard manner, such that the composition is non-associative. 

 \item DFT encodes also $\alpha'$ corrections and thereby goes beyond two-derivative approximations 
 and also beyond `generalized geometry' in that even on the half-dimensional subspace the basic 
 structures of generalized geometry are $\alpha'$-deformed.

 \end{itemize} 
In addition to the review parts, we will also present a 
number of 
new results, 
which we highlight in the following:
  \begin{itemize}
   \item We analyze the structure of {\em simultaneous} diffeomorphisms and $b$-field gauge transformations to illustrate the unusual properties of generalized
   coordinate transformations (non-standard composition and non-associativity).
   \item
   We discuss in the context of DFT the beloved chain of geometric and non-geometric backgrounds originating 
   from the constant $H$-flux on a three-torus. 
   All `gluing conditions' that make these
   spaces well-defined can be treated in a uniform manner as generalized coordinate transformations. 
   In particular, the quantization condition on the $H$-flux has a natural geometric interpretation 
   in terms of the periodicity conditions of the dual torus. Moreover, the non-geometric  
   $Q$-flux 
   background is well-defined in DFT thanks to generalized coordinate transformations that rotate 
   $x$ into $\tilde{x}$ coordinates. 
   \item We consider  
   a background that is `truly non-geometric' in the sense that it is not T-dual 
   to a geometric one. In particular, it contains $H$, $f$, and $Q$ flux simultaneously. Nevertheless,  it may be   
  globally well-defined in DFT thanks to large generalized coordinate transformations 
   that act on the fields as the relevant $O(d,d,\mathbb{Z})$ monodromies.    
   In this way, even strongly constrained DFT  
   may contain backgrounds (`T-folds') that cannot be made globally well-defined  in conventional geometry.  
  \end{itemize}

This article is organized as follows. In sec.~2 we discuss 
the generalized coordinate 
transformations (\ref{GenTRAns}) of DFT. We show how they contain conventional 
diffeomorphisms and $b$-field gauge transformations. 
The unusual composition properties of generalized coordinate transformations
arise because  
the algebra   
of infinitesimal transformations is governed 
by the Courant- or C-bracket 
rather than the Lie bracket. We make this plain 
 by studying composition 
for the explicit example of simultaneous general coordinate and $b$-field gauge transformations 
and show that, as transformations of fields, they are associative, but become non-associative 
at the level of the corresponding coordinate transformation. This is a 
form of non-associativity  
that may have a counterpart in
 closed string field theory, whose gauge algebra is of 
$L_{\infty}$ type~\cite{Zwiebach:1992ie}, rather than a strict Lie algebra. 
In section 3 we discuss the relation of $O(d,d)$ to generalized
coordinate transformations.
In sec.~4 we discuss explicit examples of non-geometric spaces 
that are not globally well-defined, but which can be patched by generalized coordinate 
transformations. For these examples the transformations take the form of 
$O(d,d)$ transformations viewed as transformations of fields but, intriguingly, do not act in the 
naive $O(d,d)$ representation on the (doubled) coordinates. 
In sec.~5 we discuss the issues related to the 
relaxation of the strong constraint. 
Finally, in sec.~6, we review 
the recently introduced $\alpha'$-extended geometry~\cite{Hohm:2013jaa},   
 which 
 requires an intriguing generalization of most geometric concepts, 
e.g., the Lie derivatives, inner product, etc. 
We close by giving a summary and outlook.

\section{Generalized coordinate transformations in DFT}
We introduce in this section the notion of generalized coordinate 
transformations in DFT and discuss 
special cases, such as conventional 
diffeomorphisms and $b$-field gauge transformations, and the subtleties of their 
geometrical interpretation. Specifically, we display   
the  type of non-associativity that emerges when  combined
diffeomorphisms and $b$-field gauge transformations 
are viewed as generalized 
coordinate transformations.

\subsection{Diffeomorphisms and $b$-field gauge transformations}
Generalized coordinate transformations $X^M \rightarrow {X'^M}$
act on tensors as in (\ref{GenTRAns}).
Thus, on the generalized metric (\ref{firstH}) we have 
  \be\label{Ftrans}
  {\cal H}_{MN}'(X') \ = \ {\cal F}_{M}{}^{K}\,{\cal F}_{N}{}^{L}\,{\cal H}_{KL}(X)\;,
 \ee
where we recall 
  \be\label{calF}
  {\cal F}_{M}{}^{N} \ = \ \frac{1}{2} \left(   \frac{\partial X^{P}}{\partial X'^{M}}\,\frac{\partial X^{\prime}_{P}}{\partial X_{N}}
  +\frac{\partial X^{\prime}_{M}}{\partial X_{P}}\, \frac{\partial X^{N}}{\partial X'^{P}}\right)\;. 
 \ee 
Let us first review how conventional general coordinate transformations are included in (\ref{Ftrans}). 
Assume that the strong constraint is solved by  
having all fields depend only on $x$, not $\tilde{x}$, 
and consider the transformation 
  \be\label{specialcoord1}
  x^{i}\;\rightarrow\; x^{i\prime} \  = \ x^{i\prime}(x)\;, \qquad \tilde{x}_{i}^{\prime} \ = \ \tilde{x}_{i}\;.
 \ee      
It can be easily seen that for this special transformation the two terms in (\ref{calF}) actually 
give the same contribution. Thus, 
  \be\label{Ftrans2}
  {\cal H}_{MN}'(X') \ = \ \frac{\partial X^{P}}{\partial X'^{M}}\,\frac{\partial X^{\prime}_{P}}{\partial X_{K}}\,
  \frac{\partial X^{Q}}{\partial X'^{N}}\,\frac{\partial X^{\prime}_{Q}}{\partial X_{L}}\,{\cal H}_{KL}(X)\;.
 \ee
Specializing to the component ${\cal H}^{ij}=g^{ij}$ and employing the usual index splitting ${V}^M=({V}_i\,,\;{V}^i)$ we obtain
  \be\label{usualgct}
  {\cal H}'^{ij}(x') \ = \ \frac{\partial \tilde{x}_{p}}{\partial \tilde{x}'_{i}}\,\frac{\partial x'^{p}}{\partial x^{k}}\,
  \frac{\partial \tilde{x}_{q}}{\partial \tilde{x}'_{j}}\,\frac{\partial x'^{q}}{\partial x^{l}}\,{\cal H}^{kl}(x)
  \ = \ \delta^i_p\,\frac{\partial x'^{p}}{\partial x^{k}}\,
  \delta^j_q\,\frac{\partial x'^{q}}{\partial x^{l}}\,{\cal H}^{kl}(x)
  \ = \ \frac{\partial x'^{i}}{\partial x^{k}}\,
  \frac{\partial x'^{j}}{\partial x^{l}}\,{\cal H}^{kl}(x)\;, 
 \ee
using in the second step that $\tilde{x}_i'=\tilde{x}_i$. This is the conventional general coordinate 
transformation of a contravariant 2-tensor ${\cal H}^{ij}$. Thus, as required, DFT correctly reproduces 
the usual diffeomorphisms acting on the (inverse) metric $g^{ij}$. Similarly, it is easy to see that 
all other components of ${\cal H}_{MN}$ transform such that they give rise to the usual diffeomorphisms 
acting on the component fields $g_{ij}$ and $b_{ij}$. 

Let us next compare with the generalized coordinate transformation of the $O(D,D)$ metric $\eta_{MN}$, 
which in components reads more precisely 
 \be
  \eta_{MN} \ = \ \begin{pmatrix}    0 & \delta^i{}_j \\[0.5ex]
  \delta_i{}^{j} & 0 \end{pmatrix}\;. 
  \ee
The transformation takes the same form as in (\ref{Ftrans2}), with ${\cal H}$ replaced by $\eta$.   
As in (\ref{usualgct}), the diagonal components $\eta^{ij}$ and $\eta_{ij}$ transform tensorially 
and therefore, being zero, they remain zero. Also the off-diagonal components transform 
tensorially.  Indeed, again from (\ref{Ftrans2}) we find 
 \be
  \eta^i{}_{j}{}^{\,\prime} \ = \ \frac{\partial \tilde{x}_{p}}{\partial \tilde{x}'_{i}}\,\frac{\partial x'^{p}}{\partial x^{k}}\,
  \frac{\partial x^{q}}{\partial x'^{j}}\,\frac{\partial \tilde{x}'_{q}}{\partial \tilde{x}_{l}}\,\eta^k{}_l
  \ = \ \,\delta^i_p \,\frac{\partial x'^{p}}{\partial x^{k}}\,
  \frac{\partial x^{q}}{\partial x'^{j}}\,\delta^l_q\,\delta^k{}_l \ = \ \frac{\partial x'^i}{\partial x^q}\,\frac{\partial x^q}{\partial x'^j}
  \ = \ \delta^i{}_j \ = \ \eta^i{}_j\;. 
 \ee 
Thus $\eta$ is invariant under general coordinate transformations. This is how it is consistent in 
DFT to have a \textit{constant} metric without restriction on the group of general coordinate 
transformations 
and therefore without restriction on allowed spacetime geometries.

Next consider the the $b$-field gauge transformations, which take the form
 \be\label{bgaugetrans}
  b_{ij}^{\prime} \ = \ b_{ij}+ 
 \partial_i\tilde{\xi}_j-\partial_j\tilde{\xi}_i\,.
 \ee
They are encoded in  (\ref{Ftrans2}), 
via the generalized coordinate transformation 
 \be\label{bgauge}
  \tilde{x}_i^{\prime} \ = \ \tilde{x}_i-\tilde{\xi}_i(x)\;, \qquad x^{i\prime} \ = \ x^i\;, 
 \ee
acting on a generalized metric that depends only on $x$.  
The details were given explicitly in section 3.1 of~\cite{Hohm:2012gk} and thus we do not 
repeat them here.
 Note that this transformation 
leaves $x$ invariant but mixes $\tilde{x}$ with $x$.

In double field theory we can also consider generalized coordinates transformations of the form
 \be\label{betagauge}
  {x'}^i \ = \ {x}^i-{\xi}^i(\tilde x)\;, \qquad \tilde x'_{i} \ = \ \tilde x_i\;, 
 \ee
i.e.~they leave $\tilde x$ invariant but mix ${x}$ with $\tilde x$. Provided the fields are now assumed to depend on $\tilde{x}$, not $x$, 
these transformations satisfy the strong constraint eq.~(\ref{STRONG}). We can parametrize
the
$O(D,D)$ matrix ${\cal H}$ in a different way using a new metric $\tilde g^{ij}$   
and a bi-vector field $\beta^{ij}$ as follows:
 \be\label{Hprime}
  {\cal H}_{MN} \ = \  \begin{pmatrix}    \tilde g^{ij} -\beta^{ik}
  \tilde g_{kl}\beta^{lj}& -\beta^{ik}\tilde g_{kj}\\[0.5ex]
  \tilde{g}_{ik}\beta^{kj} & \tilde g_{ij}\end{pmatrix} \,.    
 \ee
Then one can show that the transformation (\ref{betagauge}) acts like  a gauge transformation on $\beta$,
 \be\label{betagaugetrans}
 {\beta^{ij}}^{\prime} \ = \ \beta^{ij}+ 
 \tilde\partial^i{\xi}^j-\tilde \partial^j{\xi}^i\,, 
 \ee
and hence it is called  beta gauge transformation. As before, for certain backgrounds, the beta gauge transformations become a  $b$-field gauge transformation or
diffeomeorphisms in another T-duality frame, respectively.
We finally note that we may also evaluate the DFT action for the fields in (\ref{Hprime}), but still depending on the 
usual $x$ coordinates. In this case the action reduces to that 
discussed in relation to non-geometric fluxes in \cite{Andriot:2011uh,Andriot:2012wx,Andriot:2013xca}.

We close this subsection by discussing a form of `trivial' gauge transformations that leave the 
fields invariant. Consider a generalized coordinate transformation of the form 
 \be\label{trivialgauge}
  X'^M \ = \ X^M-\partial^M\chi\;, 
 \ee
for some function $\chi$. We view this as an exact transformation.  It was shown in 
\cite{Hohm:2012gk} that for an exact coordinate transformation of the form $X'^M=X^M-\zeta^M(X)$
the associated ${\cal F}$ can be written in terms of the matrix $a_M{}^N=\partial_M\zeta^N$ as 
\be\label{Fpert}  
{\cal F} \ = \    1 + a - a^t + \sum_{n=2}^\infty  \bigl( a^n  - 
{1\over 2} a^{n-1} a^t \bigr)   \,. 
\ee 
The strong constraint implies that $a^t a =0$, for any transformation.
Specializing to the transformation (\ref{trivialgauge}) we obtain $a_M{}^{N}=\partial_M\partial^N\chi$, 
which implies $a=a^t$ and that all powers of $a$ vanish by the strong constraint. 
It then follows from (\ref{Fpert}) that ${\cal F}=1$ and so the generalized coordinate transformation induced by 
(\ref{trivialgauge}), say on the generalized metric, reads
 \be\label{almosttrivial}
  {\cal H}'(X') \ = \ {\cal F}\,{\cal H}(X)\,{\cal F}^t \ = \ {\cal H}(X)\;.
 \ee  
We next observe that
\be
{\cal H}' (X') \ = \ {\cal H}' (X - \vec{\partial} \chi ) \ = \ {\cal H}'(X) - \partial^M \chi\,
\partial_M {\cal H}' (X)  + \cdots  \ = \    {\cal H}'(X) \,,
\ee
where we used the strong constraint that implies that all $\chi$ dependent
terms vanish.  Therefore, fields are strictly invariant under these trivial
gauge transformations: 
\be
{\cal H}'(X)\ = \ {\cal H}(X) \,. 
\ee
 An example is given by fields that are 
independent of $\tilde{x}$ and a shift $\tilde{x}_i'=\tilde{x}_i-\partial_i\chi$, 
with a function $\chi$ depending only on $x$. From (\ref{bgauge}) we identify this 
as a $b$-field gauge transformation with exact one-form gauge parameter 
$\tilde{\xi}_i=\partial_i\chi$. We thus recover the well-known `gauge symmetry of gauge symmetries'
for the $b$-fields gauge transformations. Similarly, for fields depending only on $\tilde{x}$
and a shift $x'^i=x^i-\tilde{\partial}^i\chi$, with a function $\chi$ depending only on $\tilde{x}$, 
we get a trivial $\beta$ gauge transformation (\ref{betagaugetrans}). 
Apart from these straightforward examples there
are more subtle trivial coordinate transformations for which $\chi$ may be a 
 function of both some $x$ and $\tilde{x}$. 
This redundancy in the gauge transformations of DFT will become important below, c.f.~sec.~\ref{Tfoldsection},  
where only some representative among the ``equivalent''  coordinate transformations 
is compatible with certain topological restrictions.

\subsection{Composition of generalized coordinate transformations} 
We now turn to a discussion of the general composition of the transformations (\ref{Ftrans}), 
which is different from that of ordinary diffeomorphisms. 
To explain this point it is convenient to introduce an alternative, `active' form of the transformations 
as the exponential of the infinitesimal transformations governed by generalized Lie derivatives. 
Thus, consider the transformation 
  \be\label{EXpLie}
  {V}_{M}^{\prime}(X) \ = \ \exp \big(\widehat{\cal L}_{\xi}\big) \, {V}_{M}(X)\;.
  \ee  
Note that here both sides depend on $X$, not $X'$. The question now is whether there
is an associated generalized coordinate transformation $X\rightarrow X'=X-\xi(X)+\cdots$ 
so that this implies the same field transformation as (\ref{Ftrans}). This turns out to be a 
technically rather non-trivial problem. We have confirmed in \cite{Hohm:2012gk} that, in an expansion in $\xi$, 
the two transformations agree if  
\be
X'^M \ = \  e^{-\Theta^K(\xi)\partial_K}  X^M \,, 
\ee
where 
\be~~~~
\Theta^K(\xi) \, =  \ \xi^K  \, + \,  \delta_3^K (\xi) 
+ {\cal O}(\xi^5)\,, \qquad \hbox{with} \qquad 
      \delta_3^K (\xi) \ \equiv \  {1\over 12}  \, (\xi \xi^L) \partial^K \xi_L \,, 
\ee
using the short-hand notation $\xi=\xi^P\partial_P$. 
For ordinary diffeomorphisms in standard geometry one would simply have $\Theta^M=\xi^M$.  
The extra term in $\Theta$  carries the  index on a derivative. This implies, via the strong 
constraint,  that on a scalar the action of such a transformation is like that 
of an ordinary diffeomorphism.  This is as it should be, since the generalized Lie derivative 
coincides with the ordinary Lie derivative for the case of a scalar.  
A closed form of $\Theta(\xi)$ is not known, nor the 
geometrical significance of this function. 

With the equivalent form (\ref{EXpLie})  of finite transformations we can immediately analyze composition.
Consider the consecutive action of two exponentials, with parameters $\xi_1$ and $\xi_2$, respectively,  
 \be
\label{wkdflkd}
e^{\widehat{\cal L}_{\xi_1(X)} } e^{\widehat{\cal L}_{\xi_2(X)} } \ = \ e^{\widehat{\cal L}_{\, {\xi}^c_{12}(X)} } \,.
\ee
The resulting transformation, indicated on the right hand side,  has a parameter
that can be computed with the  Baker-Campbell-Hausdorff (BCH) relation. 
Indeed, using the commutator of generalized Lie derivatives, 
 \be
  \big[\widehat{\cal L}_{\xi_1} , \widehat{\cal L}_{\xi_2} \big] \ = \  
    \widehat{\cal L}_{[\xi_1, \xi_2]_{{}_C}}\;, 
 \ee
with the C-bracket 
 \be\label{Cbracket}
   \bigl[ \xi_1,\xi_2\bigr]_{{}_C}^M  
   \ \equiv
   \ \xi_{1}^{N}\partial_{N}\xi_{2}^M -\frac{1}{2}\, 
    \xi_{1N}\partial^{M}\xi_{2}^N
   -(1\leftrightarrow 2)\;, 
 \ee
the result reads 
\be
 \label{sgxi12vgc}
 \xi_{12}^c \ = \ \xi_2 + \xi_1 + {1\over 2}  \big[\xi_2, \xi_1 \big]_{{}_C}
 + {1\over 12} \bigl(  \big[\xi_2,  \big[\xi_2, \xi_1 \big]_{{}_C} \big]_{{}_C} +  \big[\xi_1,  \big[\xi_1, \xi_2 \big]_{{}_C} \big]_{{}_C} \bigr) + \ldots\,.
 \ee
If we had the Lie bracket rather than the C-bracket in here, this formula would  encode the familiar
group structure of diffeomorphisms: acting with two diffeomorphisms is equivalent to acting with 
one that is simply the composition of maps of the first two. Here, however, we have the C-bracket, 
which in turn implies that acting with two generalized coordinate transformations is equivalent to a 
third, but this third one is not given by the direct composition of maps, thereby reflecting a novel 
group structure. 

This observation has a curious consequence because 
 the C-bracket (\ref{Cbracket}) 
has a non-trivial Jacobiator. This  
leads to a   
form of non-associativity, as we now explain.  
Acting on fields, symmetry transformations are always associative. 
Thus we must have  
 \be
\label{wkdflkd-conc-mod}
\bigl( e^{\widehat{\cal L}_{\xi_1(X)} } e^{\widehat{\cal L}_{\xi_2(X)} } \bigr)e^{\widehat{\cal L}_{\xi_3(X)} } \ = \ e^{\widehat{\cal L}_{\xi_1(X)} } \bigl( e^{\widehat{\cal L}_{\xi_2(X)} } e^{\widehat{\cal L}_{\xi_3(X)} }\bigr)  \,.
\ee
To verify this,  
we use the notation $\xi^{c}(\xi_2,\xi_1)=\xi_{12}^c$ for the parameter in (\ref{sgxi12vgc})
and note that the above    
requires  
  \be\label{EFFectivepara}
   \exp\big(\,\widehat{\cal L}_{\,\xi^c (  \xi_3 , \xi^c (\xi_2, \xi_1) )}\,\big) \ = \ 
   \exp\big(\,\widehat{\cal L}_{\,\xi^c (  \xi^c(\xi_3, \xi_2) , \xi_1  )}\,\big) \;. 
 \ee 
A straightforward computation 
shows, however, that the gauge parameters differ, 
 \be\label{Jacobiator}
\xi^c \bigl( \, \xi_3\,, \, \xi^c (\xi_2, \xi_1) \,\bigr) \ = \ \xi^c \bigl(\,  \xi^c(\xi_3, \xi_2) \,,\, \xi_1 \,\bigr)-
\frac{1}{6}J(\xi_1,\xi_2,\xi_3)+  {\cal O}(\xi^4)  \,,
\ee
where 
\be
J(\xi_1,\xi_2,\xi_3) \ = \ 
\big[\xi_1,\big[\xi_2,\xi_3\big]_{{}_C}\big]_{{}_C}+{\rm cyclic}\,,   
\ee
 is the C-bracket Jacobiator. 
This Jacobiator is actually a trivial parameter: $J^M=\partial^M N$,
where $N$ is the Nijenhuis tensor defined by 
\be\label{Nijenhuis}
 N(\xi_1,\xi_2,\xi_3) \ = \  \frac{1}{6}\Big(\big\langle \big[\xi_1,\xi_2\big]_C,\xi_3\big\rangle
  +{\rm cyclic}  \Big)\;. 
\ee  
  Generalized Lie derivatives
with trivial gauge parameters vanish. 
The effective parameters in (\ref{Jacobiator}) differ by a trivial term, 
as one can convince   
oneself that the higher order terms are also Jacobiators. As a result 
(\ref{EFFectivepara}) holds. Therefore, as field transformations the gauge 
transformations are perfectly associative. 

Intriguingly, this does not hold as transformations of coordinates. In order to see this 
consider three consecutive coordinate transformations defined by the maps $m_i$
with $i=1,2,3$: 
 \be  
\begin{split}
m_1 : \  X \to X' \,, \qquad \ \ \ \    X' \ = \ & \  e^{-\Theta(\xi_1)(X)} X\,,  \\
m_2 : \  X' \to X'' \,, \qquad  \ \   X''   \ = \ & \  e^{-\Theta(\xi_2)(X')} X'\,, \\
m_3 : \  X'' \to X''' \,, \qquad X''' \ = \ &\ e^{-\Theta(\xi_3)(X'')}X''\;.
\end{split}
\ee
The map $m_{21} = m_2\star m_1$ that defines the generalized diffeomorphism associated
with the action of $m_1$ followed by $m_2$ is given by 
\be
m_{21}\, :\quad   X \, \to\, X'' \qquad \ \ \ \    X'' \ = \ \  e^{-\Theta(\xi^c (\xi_2, \xi_1)) (X)} X \,.
\ee
Having  three successive transformations they can be implemented in two
different ways resulting in maps $m_\alpha$ and $m_\beta$:  
  \be\label{firstmap}  
  \begin{split}
m_\alpha = m_3 \star (m_2 \star m_1): \ X \to X'''_\alpha \qquad 
  X'''_\alpha \ = \ &   \ \exp\big[- \Theta(\xi^c (  \xi_3 , \xi^c (\xi_2, \xi_1) ))\big]X\;, \\
m_\beta = (m_3 \star m_2)  \star m_1: \ X \to X'''_\beta \qquad  X'''_\beta
\ = \ &   \ \exp\big[- \Theta(\xi^c (  \xi^c(\xi_3, \xi_2) , \xi_1  ))\big]X\;.  
\end{split}
 \ee  
To cubic order in parameters
 \be
 \begin{split}
 \Theta(\xi^c (  \xi_3 , \xi^c (\xi_2, \xi_1) )) \ = \ &  \ \xi^c (  \xi_3 , \xi^c (\xi_2, \xi_1) ) \ + \delta_3 (\xi_1+ \xi_2 + \xi_3)  + {\cal O}(\xi^4) \;, \\
  \Theta(\xi^c (   \xi^c(\xi_3, \xi_2) , \xi_1  )) \ =\ &  \ \xi^c (  \xi^c(\xi_3, \xi_2) , \xi_1  )) + \delta_3 (\xi_1+ \xi_2 + \xi_3)  + {\cal O}(\xi^4) \;. 
  \end{split}
 \ee
It then follows from (\ref{Jacobiator}) that
 \be\label{QJacobiator}
\Theta \bigl( \xi^c \bigl( \, \xi_3\,, \, \xi^c (\xi_2, \xi_1) \,\bigr)\bigr)  \ = \ 
\Theta \bigl( \xi^c \bigl(\,  \xi^c(\xi_3, \xi_2) \,,\, \xi_1 \,\bigr) \bigr) -
\frac{1}{6}J(\xi_1,\xi_2,\xi_3)+  {\cal O}(\xi^4)  \,.   
\ee
We therefore have 
\be
\begin{split}
X'''_\alpha (X) \ = \ & \  \exp \bigl[ \, - \, 
\Theta \bigl( \xi^c \bigl(\,  \xi^c(\xi_3, \xi_2) \,,\, \xi_1 \,\bigr) \bigr) 
+ \textstyle{1\over 6} J(\xi_1,\xi_2,\xi_3)\,+
  {\cal O}(\xi^4)  \bigr]  X \\[0.5ex]
  = \ & \  \, \exp \bigl[ \, - \, 
\Theta \bigl( \xi^c \bigl(\,  \xi^c(\xi_3, \xi_2) \,,\, \xi_1 \,\bigr) \bigr)   \bigr] 
\exp \bigl[ \,
\textstyle{1\over 6} J(\xi_1,\xi_2,\xi_3)\bigr]   X 
+ {\cal O} (\xi^4) X \\
= \ & \  \, \exp \bigl[ \, - \, 
\Theta \bigl( \xi^c \bigl(\,  \xi^c(\xi_3, \xi_2) \,,\, \xi_1 \,\bigr) \bigr)   \bigr] 
\bigl(  X  +  
\textstyle{1\over 6} J(\xi_1,\xi_2,\xi_3)  X  \bigr)    
+ {\cal O} (\xi^4) X \\
 = \ & \  X_\beta''' (X) +   
\textstyle{1\over 6} J(\xi_1,\xi_2,\xi_3)  \,  X 
+ {\cal O} (\xi^4) X \\
= \ & \  X'''_\beta (X)  + 
\textstyle{1\over 6} \vec{\partial}N (\xi_1, \xi_2, \xi_3)   
+ {\cal O} (\xi^4) X\,, 
\end{split}
\ee
where we used $J = \partial^K \hskip-2pt N \, \partial_K  = (\vec{\partial}N)^K \partial_K$.   
 In components
 we write
\be
{X_\alpha'''}^K (X) \ = \  {X_\beta '''}^K (X) + 
\textstyle{1\over 6} \partial^K \hskip-2pt N (\xi_1, \xi_2, \xi_3)  
+ {\cal O} (\xi^4) X \,. 
\ee
This is the failure of associativity of the $\star$ composition of generalized
diffeomorphisms, calculated to leading order.  Only the Jacobiator contributes
to this order.  The anomalous term $\delta_3$ in $\Theta(\xi)$ cancelled out.  
In the next subsection we will illustrate this 
fact by inspecting \textit{simultaneous} general coordinate and 
$b$-field gauge transformations.

\subsection{Simultaneous diffeomorphisms and $b$-field gauge transformations}

For definiteness we take the solution of the strong constraint for which all fields depend only on $x$. 
Then we can still consider the 
generalized coordinate transformation 
 \be\label{xprime}
   x^{i\prime} \ = \ x^{i\prime}(x)\;, \qquad \tilde{x}_i' \ = \ \tilde{x}_i-\zeta_i(x)\;, 
 \ee
corresponding to a simultaneous general coordinate and $b$-field gauge transformation. 
As the gauge parameters depend only on $x$ this is consistent with the strong constraint. 
In absence of abelian isometries these are essentially the only residual transformations compatible 
with this solution of the strong constraint. 
We compute    
  \be\label{dX1}
  \  \frac{\partial X^{\prime}_{M}}{\partial X_{P}} \ = \      
   \begin{pmatrix}   \frac{\partial x^{i \prime }}{\partial x^p} & \frac{\partial \tilde{x}'_i}{\partial x^p}\\[0.5ex]
  \frac{\partial x^{i \prime}}{\partial \tilde{x}_p} & \frac{\partial \tilde{x}^{ \prime}_i}{\partial \tilde{x}_p}\end{pmatrix}
  \ \equiv \ \begin{pmatrix}   \Lambda_p{}^{i}(x) & -\partial_p\zeta_i \\[0.5ex]
  0 & \delta^p{}_i \end{pmatrix}\;, 
 \ee
where we treated $P$ is a row index and $M$ as a column index, 
and  we introduced the notation  
 \be
   \Lambda_p{}^{i}(x)  \ \equiv \ \frac{\partial x^{i \prime }}{\partial x^p}  \;.
 \ee  
The inverse matrix is then given by 
 \be
   \frac{\partial X_{Q}}{\partial X'_{M}} \ = \      
   \begin{pmatrix}   \frac{\partial x^{q }}{\partial x^{i\prime}} & \frac{\partial \tilde{x}_q}{\partial x^{i\prime}}\\[0.5ex]
  \frac{\partial x^{q}}{\partial \tilde{x}'_i} & \frac{\partial \tilde{x}_q}{\partial \tilde{x}'_i}\end{pmatrix}\;, 
 \ee
where 
 \be\label{dX2}
   \frac{\partial x^{q }}{\partial x^{i\prime}}  \ = \ (\Lambda^{-1})_i{}^q\;, \quad
    \frac{\partial x^{q}}{\partial \tilde{x}'_i} \ = \ 0\;, \quad
    \frac{\partial \tilde{x}_q}{\partial \tilde{x}'_i} \ = \ \delta_q{}^i\;, \quad
    \frac{\partial \tilde{x}_q}{\partial x^{i\prime}} \ = \ (\Lambda^{-1})_i{}^p\, \partial_p \zeta_q\;. 
 \ee  
 With (\ref{dX1}) and (\ref{dX2}) we now can compute the various components of the matrix ${\cal F}$ defined in (\ref{calF}), 
using the usual splitting $\ {}^{M}=({}_i\;, \; {}^{i})$, 
 \be
 \begin{split}
  {\cal F}_i{}^{j} \ &= \ (\Lambda^{-1})_i{}^{j}\;, \qquad {\cal F}^i{}_{j} \ = \ \Lambda_j{}^{i}\;, \qquad {\cal F}^{ij} \ = \ 0\;, \\
  {\cal F}_{ij} \ &= \  \frac{1}{2}\Big((\Lambda^{-1})_i{}^{p}\,\Lambda_j{}^{q}\,\partial_p\zeta_q-\partial_j\zeta_i
 \  + \  (\Lambda^{-1})_i{}^{p}\big(\partial_p\zeta_j-\partial_j\zeta_p\big)\Big)\;. 
 \end{split}
 \ee
We note that in this case the two terms in ${\cal F}$ are different and thus both needed.  
As a consistency check one may verify ${\cal F}\in O(D,D)$, 
 \be
  {\cal F}_{M}{}^{P}\,{\cal F}^{N}{}_{P} \ = \ \delta_M{}^{N}\;.
 \ee  
Specializing  (\ref{Ftrans}) to components we have, for instance,  
 \be\label{gtrans} 
 \ g'^{ij}(x') \ = \   {\cal H}'^{ij}(x') \ = \ {\cal F}^{i}{}_{k}\,{\cal F}^{j}{}_{l}\,{\cal H}^{kl} \ = \ \Lambda_k{}^{i}\,\Lambda_l{}^{j}\,g^{kl}
  \ = \ \frac{\partial x^{i \prime }}{\partial x^k}\,\frac{\partial x^{j \prime }}{\partial x^l}\,g^{kl}\;, 
 \ee 
i.e., the metric still transforms with a standard general coordinate transformation. 
For the transformation of the $b$-field we have to inspect an off-diagonal component,  
 \be
 \begin{split}
 -g'^{ip}\,b'_{pj} \,  &= \,  {\cal H}'^i{}_j \   = \
  {\cal F}^{iK}\,{\cal F}_{j}{}^{L}\,{\cal H}_{KL} \ = 
  \  {\cal F}^i{}_k\,{\cal F}_{j}{}^{L}\,{\cal H}^k{}_L    \ = \ 
  {\cal F}^i{}_{k}\,{\cal F}_{j}{}^{l}\,{\cal H}^k{}_{l}+{\cal F}^i{}_{k}\,{\cal F}_{jl}\,{\cal H}^{kl} \\
  \ &= \ \Lambda_k{}^i  \Bigl[ 
  (\Lambda^{-1})_j{}^{l}\,{\cal H}^k{}_{l}
  +\frac{1}{2}
  \Big((\Lambda^{-1})_j{}^{p}\,\Lambda_l{}^{q}\,\partial_p\zeta_q-\partial_l\zeta_j
  +(\Lambda^{-1})_j{}^{p}\big(\partial_p\zeta_l-\partial_l\zeta_p\big)\Big){\cal H}^{kl}\Bigr]  \\
  \ &= -\Lambda_k{}^{i}  g^{kl} \Bigl[ (\Lambda^{-1})_j{}^p\,\,b_{lp} 
  -\textstyle{1\over 2} 
  (\Lambda^{-1})_j{}^p\,\partial_p\zeta_q\,\,\Lambda_l{}^q 
  +\frac{1}{2}\,\partial_l\zeta_j\,
  - (\Lambda^{-1})_j{}^{p}\, \partial_{[p}\zeta_{l]}  \Bigr]\,,
 \end{split} 
 \ee 
 where we use the (anti)symmetrization convention $[a b] = {1\over 2}( ab - ba)$.
Writing $g'^{ip}\,b'_{pj} =  \Lambda_k{}^i \Lambda_l{}^p g^{kl}\,b'_{pj}$
the above expression quickly yields the following transformation
for $b$: 
 \be
 \begin{split}
  b'_{ij} \ = \ &\,(\Lambda^{-1})_i{}^{k}\,(\Lambda^{-1})_j{}^{l} \,\big(b_{kl}+\partial_{[k}\,\zeta_{l]}\big) \ +\  \textstyle{1\over 2} \big((\Lambda^{-1})_{i}{}^{k}\partial_{k}\zeta_j
  -(\Lambda^{-1})_{j}{}^{k}\partial_{k}\zeta_i\big)\;. 
 \end{split}
 \ee 
In the last term 
$\Lambda^{-1} $ 
 just transforms $\partial$ into $\partial'$ and so the result can also be 
written as 
 \be
   b'_{ij} \ = \ \frac{\partial x^k}{\partial x'^i}\,\frac{\partial x^l}{\partial x'^j}\,\big(b_{kl}+\partial_{[k}\,\zeta_{l]}\big)
   +\partial'_{[i}\,\zeta^{}_{j]}\;.
 \ee  
This can be viewed as a $b$-field gauge transformation, followed by a diffeomorphism, followed 
by another $b$-field gauge transformation,  with the same parameter, but performed in the new 
coordinate system. 
An alternative writing is obtained by recalling 
that $\partial_{[k}\, \zeta_{l]}$ transforms as a two-form, so that 
we also have 
 \be\label{betterRule}
  b'_{ij}(x') \ = \ \frac{\partial x^k}{\partial x'^i}\,\frac{\partial x^l}{\partial x'^j}\,b_{kl}(x)+\frac{1}{2}\Big(\partial_i'\big(\zeta_j'(x')+\zeta_j(x)\big)
  -\partial_j'\big(\zeta_i'(x')+\zeta_i(x)\big)\Big)\;, 
 \ee 
where $\zeta_i'(x')=\frac{\partial x^k}{\partial x'^i}\zeta_{k}(x)$ is the 
coordinate transformed one-form parameter.
Thus, the resulting transformation is a general coordinate transformation together with a $b$-field gauge transformation with respect to 
a parameter that is the average of the original one and the coordinate transformed one. 
Note that for a trivial  coordinate transformation both terms agree;
in particular a trivial transformation is given by $x'^i=x^i$, $\tilde{x}'_i=\tilde{x}_i-\partial_i\chi$.

The particular combination (\ref{betterRule}) is forced on us by the original $O(D,D)$ covariant  form of 
generalized coordinate transformations, which in turn is compatible with the C-bracket. 
As we discussed  above, the C-bracket has a  
non-trivial Jacobiator, leading to the non-associativity of coordinate transformations.  
We will illustrate in the next subsection the unusual composition law
for the coordinate transformations underlying 
the transformations (\ref{betterRule}) and the non-associativity of 
successive compositions.

\subsection{Composition, Courant-bracket and non-associativity} 

Let us now ask the question how the simultaneous general coordinate and $b$-field gauge transformations 
compose. We will see that they do not compose in the naive sense of coordinate maps, i.e., 
the `group structure' will be non-trivial. 
To this end let us consider the consecutive action of two generalized coordinate transformations, 
 \be\label{twogct}
  \begin{split}
   m_1\;:\qquad\; x'^i \ &= \ x'^i(x)\;, \qquad \tilde{x}_i' \ = \ \tilde{x}_i-\zeta_{1i}(x)\;, \\
    m_2\;:\qquad x''^i \ &= \ x''^i(x')\;, \quad\;\, \tilde{x}_i'' \ = \ \tilde{x}'_i-\zeta'_{2i}(x')\;, 
  \end{split}
 \ee  
where we denoted the transformations by $m_1$ and $m_2$ for later use.  
We note that we have chosen a notation with a $'$ on $\zeta_2$ that is in principle redundant, because there is no independent definition of
a $\zeta_{2}(x)$, but it is convenient in order  to remind us with respect to which coordinate systems the parameters are 
originally defined. 
  The first transformation in (\ref{twogct}) leads by use of (\ref{betterRule}) to 
 \be\label{gentrans1}
  b'_{ij}(x') \ = \ \frac{\partial x^k}{\partial x'^i}\,\frac{\partial x^l}{\partial x'^j}\,b_{kl}(x)+\frac{1}{2}\Big(\partial_i'\big(\zeta_{1j}'(x')+\zeta_{1j}(x)\big)
  -\partial_j'\big(\zeta_{1i}'(x')+\zeta_{1i}(x)\big)\Big)\;. 
 \ee 
Similarly, the second transformations leads to 
 \be\label{gentrans2}
  b''_{ij}(x'') \ = \ \frac{\partial x'^p}{\partial x''^i}\,\frac{\partial x'^q}{\partial x''^j}\,b'_{pq}(x')+\frac{1}{2}\Big(\partial_i''\big(\zeta_{2j}''(x'')+\zeta_{2j}'(x')\big)
  -\partial_j''\big(\zeta_{2i}''(x'')+\zeta_{2i}'(x')\big)\Big)\;.
 \ee 
Inserting now (\ref{gentrans1}) in the first term in here we get 
 \be\label{STEPPP0376}
 \begin{split}
  \frac{\partial x'^p}{\partial x''^i}\,\frac{\partial x'^q}{\partial x''^j}\,b'_{pq}(x') \ 
  &= \ \frac{\partial x^k}{\partial x''^i}\,\frac{\partial x^l}{\partial x''^j}\,b_{kl}(x)
  +\frac{1}{2} \frac{\partial x'^p}{\partial x''^i}\,\frac{\partial x'^q}{\partial x''^j}\partial_p'\big(\zeta_{1q}'(x')+\zeta_{1q}(x)\big)-(i\leftrightarrow j) \\
  \ &= 
  \ \frac{\partial x^k}{\partial x''^i}\,\frac{\partial x^l}{\partial x''^j}\,b_{kl}(x) 
  +\frac{1}{2} \Bigl(  
  \partial_i''\zeta_{1j}''(x'')
  +  
  \frac{\partial x'^q}{\partial x''^j}\partial_{i}''\zeta_{1q}(x)
  -(i\leftrightarrow j)\Bigr) \;,   
 \end{split} 
 \ee
where we used again that $\partial_{[p}'\zeta_{q]}'$ transforms as a 2-form, and we used the chain rule in the last term. 
This last term can be written as 
 \be
  \frac{1}{2}\frac{\partial x'^q}{\partial x''^j}\partial_{i}''\zeta_{1q}(x) \ = \ \frac{1}{2}\partial_i''\Big(  \frac{\partial x'^q}{\partial x''^j} \zeta_{1q}(x) \Big)
  -\frac{1}{2}\frac{\partial^2 x'^q}{\partial x''^i\partial x''^j} \, \zeta_{1q} (x) \;.
 \ee 
The last term is symmetric in $i,j$ and thus drops out in (\ref{STEPPP0376}). Thus we have 
 \be
  \frac{\partial x'^p}{\partial x''^i}\,\frac{\partial x'^q}{\partial x''^j}\,b'_{pq}(x') \ = \ \frac{\partial x^k}{\partial x''^i}\,\frac{\partial x^l}{\partial x''^j}\,b_{kl}(x)
  +\frac{1}{2}\partial_i''\Big(\zeta_{1j}''(x'')+ \frac{\partial x'^q}{\partial x''^j} \zeta_{1q}(x)\Big)-(i\leftrightarrow j)\;.
 \ee 
Using this now in (\ref{gentrans2}) we obtain 
 \be\label{realComp}
  b''_{ij}(x'') \ = \  \frac{\partial x^k}{\partial x''^i}\,\frac{\partial x^l}{\partial x''^j}\,b_{kl}(x)
  +\frac{1}{2}\partial_i''\Big(\zeta_{1j}''(x'')+ \frac{\partial x'^q}{\partial x''^j} \zeta_{1q}(x)+\zeta_{2j}''(x'')+\zeta_{2j}'(x')\Big)-(i\leftrightarrow j)\;.
 \ee 
This is the final form of the consecutive action of (\ref{twogct}) on the $b$-field. 

Let us now see how this compares 
to the transformation associated with the naive
composition of the generalized coordinate
transformations.  To this end we have to view this transformation 
as a single generalized coordinate transformation $X\rightarrow X''$.  
According to (\ref{betterRule}) such a transformation acts as 
 \be\label{comptrans}
  b''_{ij}(x'') \ = \ \frac{\partial x^k}{\partial x''^i}\,\frac{\partial x^l}{\partial x''^j}\,b_{kl}(x)+\frac{1}{2} \partial_i''\Big(\zeta_{12j}''(x'')+\zeta_{12j}(x)
  \Big)-(i\leftrightarrow j)\;, 
 \ee
 where 
 \be
 \tilde x_i'' \ = \ \tilde x_i - \zeta_{12i} (x) \,, 
 \ee
for some effective parameter $\zeta_{12i}(x)$. 
Next we have to compare this parameter with the one that would emerge from direct 
composition of the two transformations (\ref{twogct}), which we denote by $\vartheta_{12}$.  
This parameter is easily computed, 
 \be
  \tilde{x}_{i}'' \ = \ \tilde{x}_i'-\zeta_{2i}'(x') \ = \ \tilde{x}_i -\zeta_{2i}'(x') -\zeta_{1i}(x) \ \equiv \ \tilde{x}_i-\vartheta_{12i}(x)\;,
 \ee 
and thus 
 \be
  \vartheta_{12i}(x) \ = \  \zeta_{2i}'(x') + \zeta_{1i}(x)\;.
 \ee 
In here, of course, we have to think of $x'$ as a function of $x$ according to (\ref{twogct}), in order for both sides to be 
functions of $x$. 
Replacing $\zeta_{12}$ by $\vartheta_{12}$ in (\ref{comptrans}) yields   
 \be
  b''_{ij}(x'') \ = \ \frac{\partial x^k}{\partial x''^i}\,\frac{\partial x^l}{\partial x''^j}\,b_{kl}(x)+\frac{1}{2} \partial_i''\Big(\zeta_{1j}''(x'')
  +\zeta_{1j}(x)+ \frac{\partial x^k}{\partial x''^j}\zeta'_{2k}(x')+\zeta_{2j}'(x')\Big)-(i\leftrightarrow j)\;.
 \ee
Comparing with (\ref{realComp}) 
it is evident that this differs, for generic values of $\zeta_1$ and $\zeta_2$, in the second term from the actual transformation $b\rightarrow b''$.  
This is, of course, what we expected, since we saw above that composition is governed by the C-bracket, which differs from the 
Lie bracket governing ordinary composition of diffeomorphisms when one considers both a non-zero 
diffeomorphism and $b$-field gauge parameter. 

It would be convenient to have a closed expression for the effective parameter $\zeta_{12}$ in terms of 
$\zeta_1$ and $\zeta_2$ in order to investigate the unconventional rules of composition in more detail. 
One can, of course, determine $\zeta_{12}$ to arbitrary order in an derivative expansion by using the 
BCH formula and the C-bracket. However, so far we did not find a simple closed expression for $\zeta_{12}$, 
but one may hope that a better analytic understanding of the composition is possible. 
It is likely that this would require a better parametrization of the transformations (\ref{twogct}).

\medskip

We close this section by 
discussing why generalized coordinate transformations of the type (\ref{twogct}) 
are necessarily non-associative as transformations of coordinates when both the diffeomorphism and 
the $b$-field gauge parameter are non-trivial. This follows from the specific form of the Jacobiator 
of the C-bracket, 
 \be
  J(U,V,W) \ = \ \big[\big[U,V\big]_{C},W\big]_{C} + \big[\big[V,W\big]_{C},U\big]_{C}+ \big[\big[W,U\big]_{C},V\big]_{C}\;,
 \ee 
which is given by 
 \be
  J(U,V,W) \ = \  \frac{1}{6}\partial^M\Big(\big\langle \big[U,V\big]_C,W\big\rangle
  +\big\langle \big[V,W\big]_C,U\big\rangle+\big\langle \big[W,U\big]_C,V\big\rangle\Big)\;, 
 \ee 
c.f.~(\ref{Nijenhuis}) above.  
From the structure of the C-bracket it follows that this vanishes for $\tilde{\partial}^i=0$ when all 
three arguments have only vector parts, i.e., when $U_i=0$, etc. In contrast, 
for non-zero vector and one-form contributions this is non-zero. In particular, the composition of
three transformations with non-zero diffeomorphism and $b$-field gauge parameter leads to 
a non-trivial Jacobiator. Thus, if we add to the two transformations $m_1$ and $m_2$ in (\ref{twogct})
a third transformation $m_3$ and if we denote the composition of generalized coordinate transformations 
by $\star$, it follows that generically  
 \be
  m_3\star(m_2\star m_1) \ \neq \ (m_3\star m_2)\star m_1\;.
 \ee 
In other words, combined diffeomorphisms and $b$-field gauge transformations are non-associative 
when viewed as coordinate transformations on the doubled space. On the contrary, as we 
explained 
above, these transformations are perfectly 
associative when acting on 
fields.\footnote{Further comments  on the diffeomorphisms of DFT can be found in \cite{Park:2013mpa}.}

\section{$O(d,d)$ as generalized coordinate transformations}  
In this section we discuss the realization of particular $O(D,D)$ transformations 
as generalized coordinate transformations of DFT. It turns out that 
in presence of $d$ commuting isometries, i.e., for field configurations that
are independent of $d$ coordinates, the $O(d,d)$ subgroup can be viewed 
as part of the generalized coordinate transformations. The action on the coordinates 
is, however, different from the naive $O(d,d)$ action but nevertheless reproduces the 
required field transformations. In order to understand this somewhat unexpected 
feature, in the first subsection we discuss again, but from a different perspective, how the 
generalized diffeomorphisms of DFT differ from ordinary diffeomorphisms. 
In the second subsection we discuss explicitly how to realize $O(d,d)$
transformations as generalized coordinate transformations.

\subsection{The role of $O(D,D)$ and its invariant metric}\label{factor2sec}
We have seen above that the presence of a gauge invariant flat metric $\eta_{MN}$ does not 
impose  
constraints on the geometry of the $D$-dimensional subspace of the doubled space since the 
gauge transformations (or coordinate transformations) are governed by generalized 
Lie derivatives (or generalized coordinate transformations).   
Here we will briefly  elucidate this point from yet another perspective.\footnote{OH thanks Axel Kleinschmidt 
for discussions on this point.}  
We explain that generalized diffeomorphisms cannot be 
viewed as ordinary diffeomorphisms in a doubled space 
subject to constraints such as, e.g., the condition that they preserve the constant metric $\eta_{MN}$.  

Naively, one may have tried to implement a doubled geometry with $O(D,D)$ metric $\eta_{MN}$ as follows. 
Start from a conventional $2D$-dimensional manifold ${\cal M}_{2D}$, thus governed by the conventional 
diffeomorphism group ${\rm Diff}({\cal M}_{2D})$, but impose the additional condition that it
respects  the  $O(D,D)$ metric $\eta_{MN}$. This would be in complete analogy to 
symplectic geometry, where one starts with a $2D$-dimensional manifold, but then imposes the 
constraint that a symplectic form is left invariant. This in turn reduces the diffeomorphism group to 
the (still infinite-dimensional) group of symplectomorphisms. Applying the 
same strategy to the  $O(D,D)$ metric we would require 
 \be
  \delta_{\xi}\eta_{MN} \ = \ {\cal L}_{\xi}\eta_{MN} \ = \ \xi^K\partial_K\eta_{MN}
  +\partial_M\xi^K\eta_{KN}+\partial_N\xi^K\eta_{KM} \ = \ 0\;.
 \ee 
Using $\eta_{MN}$ to raise and lower indices this condition becomes 
 \be\label{flatKilling}
  \partial_{M}\xi_{N}+\partial_N\xi_M \ = \ 0\;.
 \ee  
This is the usual Killing equation on a flat space, whose general 
solution is  
$\xi_M = a_M+\Lambda_{MN}X^N$, with $a_M$ and $\Lambda_{MN}=-\Lambda_{NM}$ 
constant, corresponding to translations and rigid $SO(D,D)$ transformations. 
Thus, the diffeomorphism group is reduced according to 
 \be
   {\rm Diff}({\cal M}_{2D})\;\rightarrow\; {\rm ISO}(D,D)\;.
 \ee  
If we now apply such a transformation to another tensor, say 
a vector $V_M$,   
we can use  condition (\ref{flatKilling}) and write 
 \be\label{ODDLie} 
 \begin{split}
  \delta_{\xi}{V}_{M} \ &= \ \xi^K\partial_K{V}_{M}+\partial_M\xi^K{V}_{K} \\
  \ &= \ \xi^K\partial_K{V}_{M}+\frac{1}{2}\big(\partial_M\xi^K-\partial^K\xi_M\big){V}_{K}\;. 
 \end{split}
 \ee 
This differs from the generalized Lie derivative   
$  \widehat{\cal L}_\xi {V}_{M} = \xi^K\partial_K{V}_{M}+\big(\partial_M\xi^K-\partial^K\xi_M\big){V}_{K}$ due to the  
factor of one-half in the last term. 
This shows quite clearly that the generalized diffeomorphisms cannot be 
interpreted as conventional diffeomorphisms on the doubled space. In fact, while the 
generalized Lie derivatives close according to the modified C-bracket, by construction the similarly looking 
transformations (\ref{ODDLie}) must still close according to the conventional Lie bracket.
To verify this we compute the commutator  
 \be\label{closureLie}
 \begin{split}
  \big[\delta_{\xi_1},\delta_{\xi_2}\big]V_M 
  \ = \ \,&\big[\xi_2,\xi_1\big]^K\partial_K V_M  
  +\frac{1}{2}
  \big(\partial_M\big[\xi_2,\xi_1\big]^K-\partial^K\big[\xi_2,\xi_1\big]_M\big)V_K\\
  &-\frac{1}{4}\Big(\big(\partial_M\xi_{2K}
  +\partial_K\xi_{2M}\big)\big(\partial^K\xi_1^L+\partial^L\xi_1^K\big)-
  (1\leftrightarrow 2)\Big)V_L\;, 
 \end{split} 
 \ee 
with the conventional Lie bracket 
 \be
  \big[\xi_1,\xi_2\big]^K \ = \ \xi_1^L\partial_L\xi_2^K-\xi_2^L\partial_L\xi_1^K\;.
 \ee 
The first line in (\ref{closureLie})  gives the expected transformation 
$\delta_{\xi_{12}} V_M$ with 
$\xi_{12}=[\xi_2,\xi_1]$, while in the second line only terms with the symmetrized derivative 
of the parameters  
survived, which are zero by (\ref{flatKilling}). Thus, the gauge transformations (\ref{ODDLie}) 
close according to the usual Lie bracket. 

In summary,  
if we work with a conventional $2D$-dimensional manifold with metric 
$\eta_{MN}$,  
diffeomorphisms are restricted to ${\rm ISO}(D,D)$ transformations to preserve
the metric, and their closure is governed by the usual Lie bracket. 
These transformations do not contain the 
full $D$-dimensional diffeomorphism group but only its rigid 
subgroup $GL(D,\mathbb{R})\subset {\rm ISO}(D,D)$. 
If we work instead with a {\em generalized} $2D$-dimensional manifold
 with metric $\eta_{MN}$,  we impose the strong constraint on all fields and gauge parameters.
Generalized diffeomorphisms preserve the constant form of $\eta_{MN}$
because  generalized Lie derivatives satisfy $\widehat {\cal L}_\xi \eta_{MN} =0$
 for all~$\xi^M$.  
Generalized Lie derivatives close with an algebra governed
by the C-bracket.  The strong constraint restricts the possible generalized
diffeomorphisms but 
allows arbitrary conventional 
diffeomorphisms of the $D$-dimensional subspace, thereby not posing any 
constraints on the physical spacetime.

As we will discuss below,  for backgrounds with commuting isometries, e.g., for torus backgrounds $T^d$,  
generalized diffeomorphisms 
include 
 the T-duality $O(d,d)$ transformations, 
 but with gauge parameters that differ from the naive $O(d,d)$ 
 ansatz. Specifically, in a particular form of the generalized coordinate transformations 
 (related to others by a trivial gauge transformation, c.f.~eq.~(\ref{trivialgauge}) above) it differs by factors of one-half, which are  
 related to those encountered above.

\subsection{$O(d,d)$ and dimensional reduction}\label{oddSEction}

We now discuss how for special field configurations $O(d,d)$ transformations result from 
generalized coordinate transformations. This is relevant for configurations
for which  the fields are independent of a subset of $d$ coordinates.  This condition
on the background  
holds if we have, for example, $d$ commuting isometries, taken to mean
that all fields have zero Lie derivatives along $d$ vector fields that have zero Lie brackets. 
The reduction, called strict dimensional reduction, is different from Kaluza-Klein
compatification, where massive modes arise from coordinate dependence along
the extra dimensions. 

Specifically, we split the 
coordinates into $(x^{\mu},y^{\alpha})$, $\alpha=1,\ldots,d$, and assume that metric and 
$b$-field are independent of $y^{\alpha}$. We then focus on the $O(d,d)$ action on the 
field components along the coordinate directions $y^{\alpha}$, 
which we may combine  into a $d\times d$-dimensional matrix, 
\begin{equation}
{\cal E}_{\alpha\beta} \ = \ g_{\alpha\beta}+b_{\alpha\beta}\, .
\end{equation}
The $O(d,d)$ transformations are then encoded in the usual covariant rotation of the generalized metric 
or, equivalently, by 
\begin{equation}
{\cal E}' \ = \ g_{O(d,d)} {\cal E} \ = \ \left(A{\cal E}+B\right)\left(C {\cal E}+D \right)^{-1}\, .
\end{equation}
Here $g_{O(d,d)}$ is an $O(d,d)$  group element of the form
\begin{equation}\label{GENOddmatrix}
g_{O(d,d)}=\begin{pmatrix} A & B \\ C & D \end{pmatrix}\, ,
\end{equation}
with $d$-dimensional matrices $A,B,C,D$ satisfying
\begin{equation}
A^tC+C^tA=0\, ,\quad B^tD+D^tB=0\, ,\quad A^tD+C^tB=I\, .
\end{equation}

Our task is now to realize the required $O(d,d)$  transformations as generalized 
coordinate transformations in DFT. 
Throughout this section we will not assume any topological conditions for the background configurations
and so we ignore the question whether the generalized coordinate transformations to be given below 
are `globally well-defined'. In other words, we assume the `internal' space on which $O(d,d)$
acts to be non-compact, with fields  independent of these coordinates, which is the case relevant 
for a strict dimensional reduction.
The physically more relevant case of toroidal backgrounds 
requires some care for the quantization conditions and will be discussed in detail in the next section.   

Before we  consider the relevant special subgroups of $O(d,d)$ let us first 
discuss 
the action of a generic $O(d,d)$ element $h^{M}{}_N$ viewed as the coordinate 
transformation of the (doubled) internal coordinates $Y^M=(\tilde{y}_{\alpha},y^{\alpha})$
 \be\label{ODDansatz}
  Y^{\prime M} \ = \ h^{M}{}_{N}Y^{N}\,,   ~~~\hbox{or}~~~\quad  Y' \ = \ h \, Y
    \;. 
 \ee 
A straightforward computation, whose details can be found in \cite{Hohm:2012gk}, 
shows that the above coordinate transformation 
 leads to an 
 ${\cal F}$ of the form 
 \be\label{squarerule-}
  {\cal F}^M{}_{N} \ = \
  \big[\big(h^{-1}\big)^2\big]^{M}{}_{N}\;. 
 \ee
This means that as far as  
the rotation of 
field components is concerned,  
it acts as the square of the expected matrix,  
while for the transformation of the coordinate argument it acts as expected. 
But since the fields are assumed to be independent of $Y$ this allows us to modify the 
transformations in order to fix the field transformations. We now turn to the 
various special subgroups for which we will see that the coordinate transformations can be adapted, 
e.g.~by taking the square root, 
so as to induce the required field transformation. 

\noindent
{\bf (i) $GL(d)$ transformations}\\
These are given in terms of the subgroup defined by $A,D\in GL(d)$ with 
\begin{equation}
D=(A^t)^{-1}\, ,\quad B=C=0\, .
\end{equation}
The action of 
this  $GL(d)$ subgroup of $O(d,d)$ rotates both the $y$ coordinates and their duals $\tilde y$.  
We can consider, however, the 
transformation that rotates the $y$'s but not the $\tilde y$'s:
 \be
  y' \ = \ (A^t)^{-1} y\;.
 \ee  
These are trivially contained in the generalized coordinate 
transformations of DFT.
An equivalent transformation at the level of fields, 
and thus related to the above by a `trivial' transformation, rotates $y$ and $\tilde{y}$ 
with the square root of the $O(d,d)$ matrix
(assuming this element is in the component connected to the identity).

\noindent
{\bf (ii) Shifts in the $b$-field}\\
Constant shifts in the $b$-field are given by the following matrices: 
\begin{equation}\label{bshift}
A=D=I\, ,\quad C=0 \quad \to \quad  B^t = - B\,.   
\end{equation}
These transformations can be viewed as generalized coordinated 
transformations (\ref{bgauge}) that implement 
 $b$-field gauge transformation  (\ref{bgaugetrans}), 
\begin{equation}
\tilde{y}_{\alpha}' \ = \ \tilde{y}_{\alpha}-\tilde{\xi}_{\alpha}(y)\;, \qquad \tilde \xi_{\alpha}(y) \ = \ {1\over 2}B_{\alpha\beta}y^{\beta}\, .
\end{equation}
Note that there is a additional factor of one-half
when comparing this generalized coordinate transformation with the 
$O(d,d)$ transformation induced by (\ref{bshift}) acting on the coordinates as an $O(d,d)$ vector.
This factor of one-half 
should not come as a surprise, for it corresponds precisely to the same factor 
encountered in 
sec.~\ref{factor2sec} when comparing conventional coordinate transformations on a doubled space with the generalized 
coordinate transformations of DFT. There is no contradiction in the appearance of this factor 
one-half,  
because the fields do not depend on $\tilde{y}$, and thus their
arguments are not transformed, 
with the result that the field transformations are exactly as required by the $O(d,d)$ action. 
Let us note that when considering the double torus,  
a transformations with these 
factors of one-half may be incompatible with the 
periodicity conditions. 
We will see, however, that in these cases there is an alternative 
coordinate transformation, with the same 
action on fields, i.e., related by a trivial gauge transformation, that \textit{is} 
compatible with the torus identifications.

\noindent
{\bf (iii) Shifts in $\beta$}\\
These are the transformations that are conjugate to (ii), i.e. 
\begin{equation}\label{betaODD}
A=D=I\, ,\quad B=0\quad \to \quad  C^t = - C\,.  
\end{equation}
This transformation corresponds to constants shifts of the $\beta$ field defined in (\ref{Hprime}).
It can again be viewed as a generalized coordinated transformation, this time in the form (\ref{betagauge}). 
In order to see this recall that in (\ref{squarerule-}) we obtained for the naive coordinate transformation  the 
square of the $O(d,d)$ matrix. 
Thus, while generally the $O(d,d)$ transformations cannot be viewed as generalized 
coordinate transformations, they do allow for such an interpretation in case of group elements 
connected to the identity and for backgrounds with abelian isometries 
(for which the fields do not depend on the corresponding coordinates). For then we can simply 
take the square root of the $O(d,d)$ element so that the rotation of field components is as required.  
Specialized to (\ref{betaODD}) this amounts to replacing the $O(d,d)$ matrix according to  
 \be\label{zu}
  h^{M}{}_{N}(C) \ = \ \begin{pmatrix}   \delta_{\alpha}{}^{\beta} & 0\\[0.5ex]
  C^{\alpha\beta} & \delta^{\alpha}{}_{\beta}  \end{pmatrix} \qquad \Rightarrow \qquad 
  \big(\sqrt{h}\,\big)^{M}{}_{N}(C) \ = \ \begin{pmatrix}   \delta_{\alpha}{}^{\beta} & 0\\[0.5ex]
  \frac{1}{2} C^{\alpha\beta} & \delta^{\alpha}{}_{\beta}  \end{pmatrix}\;. 
 \ee
Equivalently, this amounts to choosing  in (\ref{betagauge})
\begin{equation}\label{betaCshift}
 \xi^{\alpha}(\tilde y) \ = \  {1\over 2}C^{\alpha\beta}\tilde y_{\beta}\; , 
\end{equation}
which by construction induces the expected $\beta$ gauge transformation. Note that the parameter now depends 
on $\tilde{y}_{\alpha}$, which is compatible with the strong constraint since we assumed that there are isometries 
along the dual $y^{\alpha}$ directions. 
Note also that 
there is an additional factor of one-half when comparing this transformation with the corresponding  
element of $O(d,d)$, because we needed to take the square root. 
This, again, is perfectly consistent and to be expected in view of the discussion in sec.~\ref{factor2sec}, 
but in order to be compatible with the  torus identifications  
we will eventually adopt again 
a modified generalized coordinate transformation without factors of one-half.

\noindent
{\bf (iv) Factorized T-duality along all directions}\\ 
Finally, for 
the so-called factorized 
T-duality along all directions one may choose an $O(d,d)$ matrix of the form 
\begin{equation}
A=D=0\, ,\quad C=B=I\, , 
\end{equation}
which exchanges $x$ coordinates and $\tilde{x}$ 
coordinates.\footnote{A factorized T duality is a 
`genuine' T-duality transformation in the sense that it is
an $O(d,d)$ transformation that  does not 
belong to the `geometric subgroup' $GL(d,\mathbb{R})\ltimes \mathbb{R}^{\frac{1}{2}d(d-1)}$ that 
originates from special general coordinate transformations and $b$-field gauge transformations.}  In this case we have to distinguish between 
$d$ odd and $d$ even. In fact, for $d$ odd we are dealing with 
an $O(d,d)$ element that cannot be continuously connected with the identity element of $O(d,d)$.
In contrast, for $d$ even we can find a continuous path connecting the identity with an $O(d,d)$ transformation 
exchanging $x$ and $\tilde{x}$.  
Only in the latter case can we associate such a T-duality transformation with a generalized coordinate  transformation 
in generality, as we will now discuss, 
giving a counter-example for $d=1$. 
 
Let us now investigate the simplest example, $d=1$, in which case the most general (constant) coordinate transformation 
takes the form 
 \be
  Y' \ = \ \left( \begin{array}{c} \tilde{y}'\\ y' \\ \end{array} \right) \ = \ QY\;, \qquad Q \ = \ \begin{pmatrix}    a & b \\[0.5ex]
  c & d \end{pmatrix} \ \in \ GL(2,\mathbb{R})\;. 
 \ee
Note that for generality here we do not restrict to an $O(2,2)$ matrix, because viewed as a coordinate
transformation a priori we may employ a general, invertible transformation. 
The transformation matrix ${\cal F}$ is then given by 
 \be
  {\cal F} \ = \ \frac{1}{2}\left((Q^{-1})^t\eta Q\eta +\eta Q \eta (Q^{-1})^t\right) \ = \ 
  \frac{1}{ad-bc}\begin{pmatrix}    d^2-bc & 0 \\[0.5ex]
  0 & a^2-bc \end{pmatrix}\;. 
 \ee  
In order for this transformation to describe the overall factorized T-duality inversion it would have to be of an off-diagonal form, such as  
\be
 {\cal F}=\begin{pmatrix}    0 & 1 \\ 1 & 0 \end{pmatrix}\;.
\ee 
This follows since in a generalized coordinate transformation 
${\cal F}$ acts on the generalized metric ${\cal H}$ in the same way as an $O(d,d)$ element and so would have to 
be of the (off-diagonal) matrix form of a factorized T-duality. 
There is, however, clearly no solution for $a,b,c,d$ that satisfies this condition.

We now turn to the case that $d$ is even, focusing on $d=2$. 
For this case an explicit continuous family of $O(2,2)$ matrices connecting 
the identity with a genuine  
T-duality transformation was constructed in appendix A.3 in the second 
reference of \cite{Hohm:2011zr}. This family is defined by 
 \be\label{hfamily}
  h(\alpha) \ = \ \exp \big[ \alpha \, \big(T^{14}+T^{12}+T^{32}+T^{34}\big)  \big]\;, \qquad \alpha\in \Big[\,0,\;\frac{\pi}{2}\,\Big]\;,
 \ee 
where $(T^{MN})^{K}{}_{L}  =  2\eta^{K[M}\delta^{N]}{}_{L}$ denote the $O(d,d)$ generators in 
the fundamental representation. For $\alpha=0$ this gives the identity matrix, while for 
$\alpha=\frac{\pi}{2}$ we obtain 
  \be\label{h1h2}
 h(\tfrac{\pi}{2}) \ = \
  \begin{pmatrix}  \phantom{-}0 &\phantom{-} 0 & -1 & \phantom{-}0 \\ \phantom{-}0 &\phantom{-} 0 & \phantom{-}0 & -1 \\ -1 &\phantom{-} 0 & \phantom{-}0 &\phantom{-} 0 \\ \phantom{-}0 & -1 & \phantom{-}0 & \phantom{-}0  \end{pmatrix}\;, 
 \ee
which is the action of two T-dualities, in directions $1$ and $2$. 
Since we saw in eq.~(\ref{squarerule-}) that in order to reproduce a genuine T-duality transformation as 
a generalized coordinate transformation we have to take the \textit{square root} of the 
$O(d,d)$ matrix this is what we have to do for (\ref{h1h2}).
Since this transformations is connected to the identity 
this is straightforward and we obtain 
 \be\label{SquareRoot}
   \left(h(\tfrac{\pi}{2}) \right)^{\frac{1}{2}} \ = \ h(\tfrac{\pi}{4}) \ = \ \frac{1}{2}
     \begin{pmatrix} \phantom{-}  1 & \phantom{-} 1 & -1 & \phantom{-} 1 \\ -1 &\phantom{-} 1 &-1 & -1 \\ -1 & \phantom{-} 1 & \phantom{-} 1 & \phantom{-} 1\\-1 & -1 & -1 & \phantom{-} 1  \end{pmatrix}\;.  
  \ee 
By construction, the corresponding generalized coordinate transformation 
$X\rightarrow X'=h(\tfrac{\pi}{4})X$ induces the same $O(d,d)$ transformation \textit{on fields}
as (\ref{h1h2}).  
We will use  (\ref{SquareRoot}), in an alternative form with an identical action on fields,
to discuss a (genuinely) non-geometric background.

\section{T-folds as well-defined backgrounds in DFT}\label{Tfoldsection}
In this section we discuss particular spacetimes in DFT and their patching 
conditions as generalized coordinate transformations. 
Conventional spacetime manifolds are generally encoded in DFT via coordinate patchings that involve 
only the $x$, not the $\tilde{x}$, for which the generalized coordinate transformations 
of DFT reduce to the usual coordinate transformations of conventional differential geometry. 
In contrast, for certain special toroidal backgrounds we can allow for patching 
conditions that involve the $\tilde{x}$, in which case the coordinate  transformations are 
related  but not identical to $O(d,d)$ transformations. This will be illustrated 
with various examples.

\subsection{Generalities} \label{sec:nongeobg}
We start by discussing generalized
string backgrounds, which are configurations  that are defined 
using generalized coordinate transformations for patching together different coordinate charts instead of the standard diffeomorphisms used 
for differentiable Riemannian manifolds.   
Known examples are T-folds, 
defined in generality 
by Hull in~\cite{Hull:2004in},   
giving a framework for the novel
examples considered in~\cite{Dabholkar:2002sy,Hellerman:2002ax,Kachru:2002sk}.  
These are non-geometric string backgrounds with so-called $Q$-fluxes that are locally  
Riemannian spaces but 
fail to be so globally because 
field configurations around 
non-trivial homology cycles are glued using T-duality transformations
rather than $b$-field gauge transformations or diffeomorphisms.    
In addition to their essential role for the definition of T-folds, T-duality transformations often relate geometrical and non-geometrical backgrounds to each other.
For example,   
a three-dimensional torus with constant $H$-field is T-dual along one direction to a geometrical twisted torus, which after a T-duality transformation along a second
direction is T-dual to a three-dimensional T-fold with non-geometric $Q$-flux. Performing one more T-duality transformation along the third direction  one obtains a non-geometric $R$-flux background, leading to the well-known chain \cite{Shelton:2005cf,Dabholkar:2005ve} 
 \begin{equation}
H_{abc} \stackrel{T_{x}}{\longrightarrow} f^{a}{}_{bc}
\stackrel{T_{y}}{\longrightarrow} Q_{c}{}^{ab}
\stackrel{T_{z}}{\longrightarrow} R^{abc}\, .
\end{equation}
Here, $f^a{}_{bc} = - 2e_{[b}{}^me_{c]}{}^n\partial_me_n{}^a$ is the `geometric flux' related to the Levi-Civita spin connection. 
The $R$-flux background is usually 
thought not to be a Riemannian space, even locally.
Using DFT and dual coordinates, however, one can begin to
make sense of the background associated with
an $R$-flux space.  

The geometric interpretation of the new fluxes $Q$ and $R$, in contrast to $H$ and $f$, is not clear a priori in 
terms of the usual 10-dimensional supergravities. However, as the non-geometric fluxes are obtained through 
 $O(D,D)$ transformations we can use DFT to formulate an action for non-geometric string backgrounds, which is on equal footing with 
 the original NS action in eq.~(\ref{original}).
To be more specific, as explained in \cite{Andriot:2011uh,Andriot:2012wx,Blumenhagen:2012nk,Blumenhagen:2012nt,Blumenhagen:2013aia,Andriot:2013xca}, 
one can perform certain field redefinitions of the metric $g$ and the $b$-field, which
have the form of $O(D,D)$ transformations, and which lead to new background variables, namely a metric
 $\tilde{g}_{ij}$, a bivector $\beta^{ij}$ instead of the $b$-field, and a dilaton $\tilde{\phi}$. 
 Alternatively, we can take a different parametrization of the fundamental generalized metric 
 in terms of a metric and a bivector rather than a 2-form, see (\ref{Hprime}). 
The action written in terms of these variables  schematically takes the form
\begin{equation}\label{Qaction}
S =\int\mathrm{d}^{D}x\, \mathrm{d}^{D}\tilde{x}\,\sqrt{-\tilde{g}} e^{-2\tilde{\phi}} \,\Bigl[\, \tilde{\cal R} + 4 (\partial\tilde{\phi})^2 -\frac{1}{4} Q^2+\cdots \,\Bigr]\ ,
\end{equation}
where the dots represent terms that vanish
when we set the differential operators 
$\beta^{ij}\partial_j$ and $\tilde{\partial}^i$ equal to zero. 
The non-geometric $Q$-flux 
(replacing $H$) is given by 
\begin{equation}
\label{eq:qflux}
  Q_{i}{}^{jk} \ = \ \partial_{i}\beta^{jk}\;.
 \end{equation}
 This action is  invariant under beta gauge transformations of the form (\ref{betagaugetrans}).
 Although this is not manifest in the above 
 form, one can show that $Q$ plays a natural role as (part of) a
 new connection $\tilde{\nabla}^i$
 for the winding derivatives 
 \cite{Andriot:2012wx}, see also \cite{Andriot:2013xca}. 
 More precisely, the $Q^2$ term in  the above action 
 then becomes part of a second (dual) Einstein-Hilbert term involving 
 winding derivatives. 
  Moreover, as discussed in \cite{Andriot:2012wx}, the full DFT action written in this form also encodes 
$R$-flux contributions, which read
 \be
  R^{ijk} \ = \ 3\big(\tilde{\partial}^{[i}\beta^{jk]}+\beta^{p[i}\partial_p\beta^{jk]}\big)\;. 
 \ee
This is a tensor under the `beta gauge transformations' (\ref{betagaugetrans}) parametrized by $\xi^i$. 
Its leading term is the complete T-dual of the $H$-flux and not visible in conventional geometry with only $x$ coordinates. 
Its subleading term, however, is visible in conventional supergravity written in terms of the new variables $\tilde{g}_{ij}$ and $\beta^{ij}$.   
We finally note that one can also show  \cite{Hassler:2013wsa}  that non-geometric objects like $Q$-branes are globally well-defined solutions of the action (\ref{Qaction}), in analogy to their T-dual counterparts, the NS-5-branes,
which are well-defined  
solutions of the original action (\ref{original}). 

In the following  we will  discuss some aspects of non-geometric backgrounds and their corresponding $O(D,D)$ monodromy transformations.
We will show that  
these can be
seen as generalized coordinate transformations, which define the generalized patching conditions of the non-geometric background spaces.
In particular, we will discuss a background that is genuinely non-geometric in the sense that it is 
not T-dual to a geometric background, 
but which can nevertheless be consistently defined in DFT by virtue of generalized coordinate 
transformations that take the form of factorized T-dualities.

The $D=(d+d')$-dimensional backgrounds ${\cal M}^{d+d'}$ to be considered in the following can be described in a convenient uniform manner: they all take the form, at least locally, of a fibration of a $d$-dimensional  torus $T_f^d$  over a $d'$-dimensional 
base ${\cal B}^{d'}$:  
\begin{equation}\label{fibration}
T^d_{f}\,\hookrightarrow\, {\cal M}^{d+d'}\,\rightarrow\, {\cal B}^{d'}\, .
\end{equation}
In our examples, the base space will be the one-dimensional circle $S^1$.

\begin{figure}[t]
\centerline{\epsfig{figure=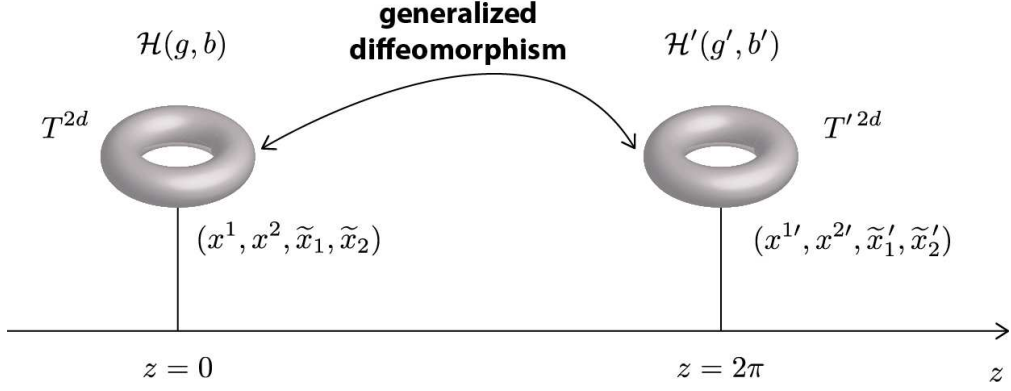, height=5.1cm}}
\caption{Construction of a background field configuration for a double-torus
fibration over a circle $z \sim z+ 2\pi$. A generalized diffeomorphism of the
double torus induces the desired identifications of fields
at $z=0$ and $z=2\pi$.}
\label{f-Z2}
\end{figure}

We will show that all backgrounds to be defined below are globally well-defined in DFT according to 
the following universal picture. The coordinates are split into $x^1, x^2$ for the 2-torus (augmented 
by the dual coordinates $\tilde{x}_1, \tilde{x}_2$ for the doubled torus) and the coordinate $z$, 
with the identification $z\sim z+2\pi$. 
We consider backgrounds whose metric and $b$-field depend only on $z$. In order to 
show that such a background is globally well-defined we have to verify that the metric and 
$b$-field at $z=0$ and $z=2\pi$ are gauge equivalent and so can be consistently `glued together'. 
In standard supergravity this is the case if they are related by a diffeomorphism or a $b$-field 
gauge transformation. However, DFT also allows for genuinely non-geometric backgrounds, for 
which generalized coordinate transformations are required that take the form of genuine T-duality 
transformations. More specifically, we will show that in each case there is a generalized coordinate 
transformation of the doubled torus coordinates $(x^1,x^2,\tilde{x}_1,\tilde{x}_2)$ so that 
 \be
  {\cal H}'(g',b')(z=2\pi)  \ = \    {\cal H}(g,b)(z=0) \;, 
 \ee
as depicted graphically in Figure~\ref{f-Z2}. 

While for conventional backgrounds the generalized diffeomorphism used in the gluing acts 
only on $(x^1, x^2)$ and so reduces to a conventional diffeomorphism, for the non-geometric backgrounds 
a generalized diffeomorphism is required that acts non-trivially on the full doubled coordinates.   
Before we discuss this in detail, we review in the next subsection the relation between the 
needed $O(d,d)$ transformations and generalized diffeomorphisms.

We now illustrate the above general results with various examples motivated by non-geometric string compactifications 
and non-geometric fluxes. 
For comparison with the literature on the subject it is instructive to give these transformations 
also in terms of the K\"ahler parameter 
$\rho$ and the complex structure $\tau$ of the torus, which are given by 
 \be\label{rhoandtau}
  \rho \ = \ -b_{12}+i\ V\;, \qquad \tau \ = \ {\frac{g_{12}}{g_{11}}}+i\ {\frac{V}{g_{11}}}\;, 
 \ee
where $V$ is the volume of the 2-torus.  
The T-duality group $O(2,2)$ acts on $\rho$ and $\tau$ according to the  isomorphism
 \be\label{so22iso}
  SO(2,2) \ \cong  \ SL(2)_{\tau}\times SL(2)_{\rho}\;, 
 \ee
with the usual $SL(2)$ action on $\tau$ and $\rho$
 \be
   \tau' \ = \ \frac{a\tau+b}{c\tau+d}\;, \qquad \rho' \ = \ \frac{a'\rho+b'}{c'\rho+d'}\;. 
 \ee
 The explicit embedding for the $SL(2)$ parameters into an $O(D,D)$ matrix as (\ref{GENOddmatrix}) reads 
 \cite{Condeescu:2013yma}, 
 \be
  A \ = \  a'\begin{pmatrix}a&b\\ c&d \end{pmatrix}\,, \quad  B \ = \  b'\begin{pmatrix}-b&a\\ -d&c \end{pmatrix}\,,\quad  C \ = \  c'\begin{pmatrix}-c&-d\\ a&b \end{pmatrix}\,,
  \quad  D \ = \  d'\begin{pmatrix}d&-c\\ -b&a \end{pmatrix}\,. 
 \ee

\subsection{3-torus with constant $H$-flux}\label{Hfluxsec}

We first consider a three-dimensional target space, which we take to be a flat torus along the directions $x^{i}=(x^1, x^2, x^3 =z)$, with an $H$-flux $H_3=\bar H d x^1 
\wedge d x^2 \wedge d z$, where $\bar H$ is a constant. 
The $H$ flux is quantized   
according to \cite{Kachru:2002sk}\footnote{This holds for conventions in which the world-sheet string action is given by 
\be\nonumber
 S = -\frac{1}{4\pi\alpha'}\int d\sigma d\tau\left(\sqrt{h}h^{\alpha\beta}\partial_{\alpha}X^i\partial_{\beta}X^j\,g_{ij}
 +\epsilon^{\alpha\beta}\partial_{\alpha}X^i\partial_{\beta}X^j\, b_{ij}\right)\;. 
\ee}
 \be\label{fluxquant}
  \frac{1}{(2\pi)^2\alpha'}\int H_3  \ \in \ \mathbb{Z}\;.
 \ee
For now we use conventions for which the length dimensions of fields, constants  and coordinates   
is given by \be
  [g] \ = \ [b] \ = \ 0\;, \quad [\alpha'] \ = \ L^2\;, \quad [x] \ = \ L\;,  \quad [\bar H] \ = \ \frac{1}{L} \;,
 \ee
where $x$ collectively denotes the coordinates $x^i$ and $L$ is a length scale. 
The metric takes the constant form $g_{ij}=\delta_{ij}$ and the coordinates are identified 
according to $x^i\sim x^i+2\pi R_i$.
Therefore the flux integral in (\ref{fluxquant}) is given by 
 \be
  \int H_3 \ = \ \int \bar H d^3 x \ = \ (2\pi)^3 R_1 R_2 R_3\, \bar H\;, 
 \ee 
so that the quantization condition (\ref{fluxquant}) becomes 
 \be\label{2ndquantization}
  \frac{2\pi R_1R_2R_3\, \bar H}{\alpha'} \ \in \ \mathbb{Z}\;.
 \ee
The $b$-field, in a particular gauge, can be written as 
 \be
  b \ \equiv \ b_{12} \ = \ -b_{21} \ = \ \bar H\,z\;, 
 \ee
with all other components set to zero.  
 
 Next, we investigate the question whether this background is globally well-defined. 
To this end we have to compare the field configurations  
 at $z=0$ and $z=2\pi R_3$.
While $g$ is constant, we find that for $b$
 \be
  b(2\pi R_3)-b(0) \ = \ 2\pi R_3 \bar H\;.
 \ee 
This can be compensated by a $b$-field gauge transformation
acting on the fields living in a neighborhood of  $z= 2\pi R_3$.  
A possible choice is:
\begin{equation}\label{bfieldGauge}
\tilde{\xi_1} \ = \  2\pi R_3 \bar Hx^2\,, \;\; \tilde{\xi}_2 \ = \ 0 \;\,
\quad\to  
\quad\quad  b' \ = \  b -2\pi R_3 \bar H\, , 
\end{equation}
so that after this gauge transformation
 \be
  b'(2\pi R_3) \ = \ b(2\pi R_3)-2\pi R_3 \bar H \ = \ b(0)\;,
 \ee
and the space is globally well-defined despite the apparent lack of periodicity in $z$.

As reviewed earlier, the 
above $b$-field gauge transformation can also be realized as a  generalized coordinate transformation 
which, noting (\ref{bfieldGauge}),
is written as
\bea\label{htorus}
x^{1\prime} &= &x^{1}  \; ,\nonumber\\
x^{2\prime} &= &x^{2}  \; ,\nonumber\\
\tilde x_1^{\prime} &=& \tilde x_{1} -2\pi R_3 \bar H  x^{2}\;, \\
\tilde x_2^{\prime} &= &\tilde  x_{2} \;  .   \nonumber
\eea
Therefore this background possesses patching conditions that are naturally viewed as DFT generalized coordinate 
transformations, mixing $x$ and $\tilde{x}$ coordinates and thus treating the required general coordinate and  $b$-field gauge transformations on the same footing.

Let us now show 
that the above coordinate transformations on the double
torus in fact require the quantization of the $H$-flux.  For this we take
the $\tilde x_1$ circle to have the radius $\tilde R_1$ given by T-duality:
\be\label{dualradius}
  \tilde{R}_1 \ = \ \frac{\alpha'}{R_1}\;.   
 \ee
Then, in (\ref{htorus}), if we shift $x^2\rightarrow x^2+2\pi  R_2$ we must get
the same point on the dual torus. This requires that 
the resulting shift in $\tilde x_1$ be some integer multiple
 $n$ of $2\pi \tilde R_1$:
\be
(2\pi R_3) (2\pi R_2) \,\bar H  \ = \ n \, 2\pi \frac{\alpha'}{R_1}\quad \to \quad
2\pi \, R_1 R_2 R_3\, {1\over \alpha'} \, \bar H  \ = \ n \,,
\ee 
which is precisely the quantization condition (\ref{2ndquantization}) of the flux.
We could have realized the same duality transformation by the coordinate 
transformation $\tilde{x}_2' = \tilde{x}_2+2\pi R_3 \bar H x^1$, which by the same argument gives the flux quantization upon  taking
the dual circle along $\tilde{x}_2$ to have radius 
$\tilde{R}_2=\frac{\alpha'}{R_2}$. 
We have thus arrived at a purely geometric 
perspective on the flux quantization condition.  
Let us note, finally, that for this background DFT is 
not strictly needed as the patching may also be done in conventional 
supergravity using the allowed $b$-field gauge transformations. This will change for some examples below.

We next discuss the above example for slightly different conventions, 
which will be more convenient below. Specifically, we will work with dimensionless coordinates so that, e.g., 
the metric becomes dimensionful, and the radii no longer enter the coordinates but the metric. 
To this end we perform the coordinate transformation $x^i\rightarrow \hat{x}^i=\frac{x^i}{R_i}\in[0,2\pi]$, 
dropping the hats shortly. 
This leads to a metric that is given in terms of the radii $R_{i}$ by
\begin{equation} \label{Ametric}
g \ = \ \begin{pmatrix} R_1^2 & 0 & 0 \\ 0 & R_2^2 & 0 \\ 0 & 0 & R_3^2 \end{pmatrix} \ .
\end{equation}
The new $b$-field after this coordinate transformation ($\hat b \equiv \hat b_{12}$)
 is given by 
 \be
  \hat{b} \ = \ R_1 R_2\, b \ = \ R_1R_2\, \bar H\,z \ = \ R_1R_2R_3 \,
  \bar H\,\hat{z} \ \equiv \ \alpha' H\,\hat{z}\;, 
 \ee 
where we introduced 
 \be
 \label{trhz}
  {H} \ \equiv \ \frac{R_2R_2R_3 \bar H}{\alpha'} \quad \to \quad 2\pi H\in\mathbb{Z}\,,
 \ee
and rewrote the quantization condition (\ref{2ndquantization}).   
In these conventions the $\hat b$-field has dimension $L^2$ and so has its one-form gauge parameter $\hat{\tilde{\xi}}$, since the hatted coordinates are dimensionless.
We have that, schematically, 
 $\hat{\tilde{\xi}}=R\tilde{\xi}$, which leads to the expected form for the $b$-field gauge transformations
$\delta \hat{b}=\hat{\partial}\hat{\tilde{\xi}}$. 
 This leads to an $\alpha'$ dependence in the relation 
between $b$-field gauge transformations and generalized coordinate transformations. In fact, 
defining dimension-free dual coordinates with standard
periodicity $\hat{\tilde{x}}\equiv\frac{\tilde{x}}{\tilde{R}}=\frac{R}{\alpha'}\tilde{x}$, 
makes the coordinate transformation $\tilde x' = \tilde x - \tilde \xi$ turn into 
 \be
  \hat{\tilde{x}}' \ = \ \hat{\tilde{x}}-\frac{R}{\alpha'}\tilde{\xi} \ = \  \hat{\tilde{x}}-\frac{1}{\alpha'}\hat{\tilde{\xi}}\;.
 \ee
Thus, working with dimensionless coordinates and metric and $b$-field of length dimension $L^2$ 
we have a relative factor of $\alpha'$ between the $b$-field gauge transformation and the generalized 
diffeomorphism.   
Dropping the hats on fields and coordinates 
and setting $\alpha'=1$, the 
background field values are given by (\ref{Ametric}) and a $b$-field that is, in a particular gauge, linear in $z$,
\begin{equation} \label{Abfield}
b \ = \ b_{12} \ = \ - b_{21} \ = \  H z  \; ,
\end{equation}
while the periodicity condition of the double torus is 
$x^{i}\sim x^{i}+2\pi$ and $\tilde x_{i}\sim \tilde x_{i}+2\pi$. This is the convention we use from now on.

  We close this section by mentioning  that one can also view the above transformations as $O(2,2)$ monodromy 
transformations in the 1,2-directions.
We first note that  the $b$-field gauge transformation (\ref{bfieldGauge})  acts on the K\"ahler parameter as 
\be\label{rhoshift}
\rho' \ = \  \rho +  2\pi R_3 \bar H \, , 
\ee
while $\tau$ is invariant. With the above isomorphism this corresponds to the 
$O(2,2)$ matrix
\be\label{bfieldodd}
h \ = \ 
\begin{pmatrix}1&0&0& 2\pi R_3 \bar H\\ 0&1&-2\pi R_3 \bar H&0\\0&0&1&0\\0&0&0&1 \end{pmatrix}\  \, .
\ee
In particular, it is straightforward to verify (\ref{rhoshift}) and the invariance of $\tau$.

\subsection{Twisted 3-torus: $f$-flux}\label{subsubsec:Ttorus}
Now we consider a twisted 3-torus that is a field configuration related to the flat torus with $H$-flux discussed above 
by a T-duality in the 1-direction \cite{Kachru:2002sk}. 
The metric is independent of $x^{\alpha}$, $\alpha=1,2$, but depends on the coordinate $z$, 
\begin{equation}
g  \ = \ \begin{pmatrix} \frac{1}{R_1^2} & - \frac{Hz}{R_1^2} & 0 \\
\phantom{\Biggl(}\hskip-10pt  - \frac{Hz}{R_1^2} & R_2^2 + \bigl(\frac{Hz}{R_1}\bigr)^2 & 0 \\[0.3ex] 0 & 0 & R_3^2 \end{pmatrix} \ .\label{metric}
\end{equation}
This is a three-dimensional so-called nilmanifold 
with no additional  $b$-field. 

We want to investigate whether this background is globally well-defined. 
As above we have to compare the metric at $z=0$ and $z=2\pi$. 
One finds after a quick computation for the complex structure $\tau$ of the 2-torus defined in (\ref{rhoandtau}) 
\be
\tau(2\pi) - \tau(0) \ = \ - 2\pi H \, ,
\ee
while $\rho$ is unchanged. 
This lack of periodicity  can be compensated by an $O(2,2)$ transformation in the 1,2-directions, 
so that $\tau'(2\pi)=\tau(0)$, 
with the following matrix 
\be\label{O22Hsecond}
h \ = \ 
\begin{pmatrix}1&2\pi H&0&0\\ 0&1&0&0\\0&0&1&0\\0&0&-2\pi H&1 \end{pmatrix}\;, 
\ee
where we used again the explicit isomorphism (\ref{so22iso}).
This matrix belongs to the $GL(2)$ subgroup. Thus, 
we can can reproduce this transformation 
by an accompanying general coordinate transformation in the 
$(x^1,x^2)$  coordinates, 
\bea
x^{1\prime} &= &x^{1} \; ,\nonumber\\
x^{2\prime} &= &x^{2}-2\pi Hx^1  \; ,\nonumber\\
\tilde x_1^{\prime} &=& \tilde x_{1} \; ,\nonumber\\
\tilde x_2^{\prime} &= &\tilde  x_{2}  \;. 
\label{eq:bdyXmonod1}
\eea
Thus, the 
space is globally well-defined. Note that the $\tilde{x}$ coordinates are not transformed, 
in contrast to the naive $O(2,2)$ action of (\ref{O22Hsecond}) on $X^M$. 
We could have realized the same field transformations with a (dual) transformation on 
$(\tilde{x}_1,\tilde{x}_2)$, leaving $(x^1, x^2)$ invariant.\footnote{The square root of (\ref{O22Hsecond}), acting both 
on the $x$ and $\tilde{x}$ coordinates, also leads to the same field transformation but is illegal 
in general due to the periodicity implied by the torus topology.}

\subsection{3-torus with $Q$-flux: T-fold}\label{sec:nongeobg}

Finally, we consider the situation
obtained from the previous example 
by a T-duality transformation
in the 2-direction. This gives the background
\begin{equation}
g \  = \ f \begin{pmatrix}  \frac{1}{R_1^2} & 0 & 0 \\ 0 & \frac{1}{R_2^2} & 0 \\ 0 & 0 & \frac{R_3^2}{f} \end{pmatrix}\ ,\qquad 
 \ b \ = \ f \begin{pmatrix} 0 & -\frac{Hz}{R_1^2 R_2^2} & 0 \\ \frac{Hz}{R_1^2 R_2^2} & 0 & 0 \\ 0 & 0 & 0 \end{pmatrix}\; , \label{eq:nongeofields}
\end{equation}
where 
 \be
  f (z) \ = \ \Big[1+\Bigl(\frac{H z}{R_1R_2} \Bigr)^2 \, \Big]^{-1} \;. 
 \ee 
This configuration is known to be non-geometric. Indeed, as we will now discuss, when going around the $z$ circle   
one cannot find a general coordinate or 
$b$-field gauge transformation 
which would make these fields globally well-defined.
We can achieve this, however,  by using a generalized coordinate transformation
corresponding to
a $\beta$-shift as the transition function between two patches on the $z$ circle.

Let us compute the field values at $z=0$ and $z=2\pi$ for 
the K\"ahler parameter (\ref{rhoandtau}) of the 2-torus. 
We first compute with (\ref{eq:nongeofields}) that 
 \be
  \rho(z) \ = \ \frac{1}{Hz-iR_1R_2}\;.
 \ee
This implies   
\be\label{nonlinrho}
\rho(2\pi) \ = \ \frac{1}{2\pi H-iR_1R_2} \ = \ \frac{1}{2\pi H+\frac{1}{\rho(0)}} \ = \ \frac{\rho(0)}{1+2\pi H \rho(0)} \; .
\ee
This (non-linear) transformation on $\rho$ cannot be viewed as 
a gauge symmetry of supergravity. It  can be viewed, however, as an $O(2,2)$ transformation in the 1,2-directions. 
Thus it can be compensated with the inverse transformation, given by the following matrix 
\be\label{O22Hthird}
h \ = \ 
\begin{pmatrix}1&0&0&0\\ 0&1&0&0\\0& 2\pi H&1&0\\-2\pi H&0&0&1 \end{pmatrix}\; ,
\ee
as follows again with (\ref{so22iso}). 
Thus, in order to make this space globally well-defined one needs to resort 
to some notion as T-fold, where one allows for `patchings with $O(d,d)$ transformations'.
In standard supergravity, where $O(d,d)$ cannot be viewed as part of the 
gauge group, this is little more than words, but in DFT this idea can be 
given a concrete meaning.  
In fact, as  reviewed above, $O(2,2)$ transformations can be viewed as 
generalized coordinate transformations in DFT.
Since (\ref{O22Hthird}) takes the form of a $\beta$ transformation 
we infer from sec.~\ref{oddSEction}(iii) that we need the generalized 
coordinate transformation (\ref{betaCshift}) to reproduce the non-linear 
transformation (\ref{nonlinrho}) of $\rho$. 
One obtains 
\bea\label{firstattempt}
x^{1\prime} &= &x^{1} +\pi H\tilde x_2 \; ,\nonumber\\
x^{2\prime} &= &x^{2}-\pi H \tilde x_1  \; ,\nonumber\\
\tilde x_1^{\prime} &=& \tilde x_{1} \; ,\nonumber\\
\tilde x_2^{\prime} &= &\tilde  x_{2} \;  .
\eea
However, due to the factors of $\pi H$ rather than $2\pi H \in \mathbb{Z}$, this transformation is not compatible 
with the torus identifications, 
but we can consider an alternative 
transformation, acting in the same way on fields, that 
rotates $x^1$ and $x^2$ 
as follows  
 \bea
x^{1\prime} &= &x^{1} +2\pi H\tilde x_2 \; ,\nonumber\\
x^{2\prime} &= &x^{2}   \; .
\eea
This transformation is well defined and implements the desired transformation
of fields. It  differs from (\ref{firstattempt}) by a trivial gauge transformation of the 
type (\ref{trivialgauge}), with 
 \be
  \chi \ = \ -\pi H \tilde{x}_1\tilde{x}_2\;. 
 \ee 
This shows that in DFT 
this space is globally well-defined, 
given the generalized coordinate transformations mixing $x$ and $\tilde{x}$ 
used for the 
patching after going around  the circle in $z$ direction. 
We finally note that the parameter $H$ is now associated with a non-geometric $Q$-flux.

\subsection{Genuine non-geometric backgrounds}

After discussing the above chain of backgrounds obtained from the geometric $H$-flux background by T-duality transformations 
we now want to consider  a background that is genuinely non-geometric in the sense that it 
is not T-dual to a geometric background. (Such backgrounds have also been discussed in \cite{Schulgin:2008fv}.) 
As discussed in \cite{Condeescu:2012sp,Condeescu:2013yma}
this background can be constructed as a truly asymmetric $\mathbb{Z}_2$ orbifold CFT.
The corresponding background is a fibered torus with the following complex structure and K\"ahler parameters:
\begin{eqnarray}\label{nongeometaurho}
\tau(z)&=&{\tau_0\cos (fz)+\sin (fz)\over\cos (fz)-\tau_0\sin (fz)}\, ,\quad \ \ \,  f\in{1\over 4}+{\mathbb Z}\, ,\nonumber\\
\rho(z)&=&{\rho_0\cos (Hz)+\sin (Hz)\over\cos (Hz)-\rho_0\sin (Hz)}\, ,\quad H\in{1\over 4}+{\mathbb Z}\, .
\end{eqnarray}
Here $\tau_0$ and $\rho_0$ are arbitrary parameters of the background.
In fact, this three-dimensional background does not solve the beta function equations for
the underlying non-linear sigma model for arbitrary parameters $\tau_0$ and $\rho_0$; 
it is not a conformal field theory. 
Only for  
$\tau_0=\rho_0= i$, the background becomes an CFT,
namely the exactly solvable, freely acting asymmetric $\mathbb{Z}_2$ orbifold CFT.
However, we may still consider these backgrounds as off-shell configurations for arbitrary parameters. 
One can show that this background is not T-dual to a geometric background. 
We also mention that 
this  background contains $f,H$ and $Q$ flux. Specifically one can  identify the geometric flux with the $f$-parameter, 
whereas $H, Q$ are generated by $H$.

In order to analyze the global structure we compare again the fields at $z=0$ and $z=2\pi$. We find 
an inversion of the complex structure as well as an inversion of the K\"ahler parameter of the fibre torus:
\be\label{taurho}
\tau(2\pi) \ = \  -{1\over \tau(0)}\, , \qquad\rho(2\pi) \ = \ -{1\over \rho(0)}\, .
\ee
The fixed point of this transformation, $\tau_0=\rho_0= i$, 
precisely agrees with the asymmetric orbifold point, mentioned already above.
Note that at the fixed point this transformation (\ref{taurho}) has trivial action on $\tau$ and on $\rho$.
Nevertheless, it corresponds to an $O(2,2)$ monodromy transformation, which will still act non-trivially 
on the coordinates and the dual coordinates. 
The transformation (\ref{taurho}) corresponds to the following $O(2,2)$ monodromy transformation:
\be\label{finalmonodromy}
h \ = \ 
\begin{pmatrix}
\phantom{-}0&\phantom{-}0&-1&\phantom{-}0\\ \phantom{-}0&
\phantom{-}0&\phantom{-}0&-1\\-1&\phantom{-}0&\phantom{-}0&
\phantom{-}0\\\phantom{-}0&-1&\phantom{-}0&\phantom{-}0 \end{pmatrix}\,  .
\ee
As discussed in sec.~\ref{oddSEction}(iv), see (\ref{h1h2}), this particular $O(2,2)$ group element, which corresponds to an overall T-duality transformation,  
can be associated to a generalized coordinate transformation that is determined by the square root (\ref{SquareRoot}) 
of this matrix.  However, since the matrix entries of  (\ref{SquareRoot}) are half-integer valued this 
transformation is not compatible with the torus 
identifications 
assumed for the (doubled) coordinates. 
In order to remedy this we can give an alternative form of the generalized coordinate transformation that 
results from this one by a trivial gauge transformation, thus giving the same transformation of fields, 
however, in a way that is compatible with the torus identifications.   

We now give the details of this `trivial' generalized coordinate transformation. The original (illegal) 
transformation we write as  
 \be\label{constantcoord}
  X'^M \ = \ A^{M}{}_{N}\,X^N\;, 
 \ee
where $A$ is the constant matrix (\ref{SquareRoot}). 
The induced generalized coordinate transformation on fields is determined by the 
associated ${\cal F}$, which reads 
 \be\label{assocF}
  {\cal F}(A)^M{}_{N} \ = \ \frac{1}{2}\left(\eta(A^{-1})^{T}\eta A+A\eta(A^{-1})^T\eta\right)^M{}_{N}\;. 
 \ee
We now act with a second, trivial transformation 
 \be
  X''^{M} \ = \ X'^M+\partial'^M\chi \ = \ A^{M}{}_N\,X^N+(A^{-1})_N{}^{M}\,\partial^N\chi\;, 
 \ee
where we used the chain rule in the second equation (and indices 
on $A^{-1}$ are raised and lowered with $\eta$). Thus, in matrix notation we have 
 \be\label{trivialXprime}
  X'' \ = \ AX+\eta(A^{-1})^T\eta\,\vec{\partial}\chi \ = \ A\left(X+ \vec{\partial}\chi\right)\;, 
 \ee  
where we used $A\in O(2,2)$, which implies $A=\eta(A^{-1})^T\eta$.   
We now give a function $\chi$ that leads to a particularly simple coordinate transformation 
that induces the required field transformation. It reads\footnote{Note that 
$\chi$ does not satisfy the weak constraint $\partial^M \partial_M \chi =0$. 
While perhaps surprising, $\chi$ still gives a trivial gauge transformation
because this only requires $\partial^M \chi\, \partial_M A =0$, for any field
$A$. This condition is satisfied because no field $A$ depends on $x^1, x^2$
nor $\tilde x_1 , \tilde x_2$. } 
 \be
 \begin{split}
  \chi(x^1,x^2,\tilde{x}_1,\tilde{x}_2) \ = \ \,&\frac{1}{2}\left(x^1x^2+\tilde{x}_1\tilde{x}_2-\tilde{x}_1x^2-x^1\tilde{x}_2\right)
  -\frac{3}{2}\left(\tilde{x}_1x^1+\tilde{x}_2 x^2\right)\\
  &-\frac{1}{4}\left((x^1)^2+(x^2)^2+(\tilde{x}_1)^2+(\tilde{x}_2)^2\right)\;. 
 \end{split} 
 \ee  
Inserting (\ref{SquareRoot}) for $A$ and this ansatz for $\chi$ in (\ref{trivialXprime}) 
it is straightforward to compute the resulting coordinate transformation $X\rightarrow X''$. 
Viewing this as a single coordinate transformation it is more convenient to write 
the new coordinates as $X'$, and we finally get 
 \be
 \begin{split}
  \tilde{x}_1' \ &= \ -\tilde{x}_2\;, \\
  \tilde{x}_2' \ &= \  \ \,  x^1\;, \\
  x'^1 \ &= \ -x^2\;, \\
  x'^2 \ &= \ \ \, \tilde{x}_1\;.
 \end{split}
 \ee 
This corresponds to a constant coordinate transformation of the form (\ref{constantcoord}), 
with matrix 
 \be
  A \ = \ 
  \begin{pmatrix}0&-1&0&0\\ 0&0&1&0\\0&0&0&-1\\1&0&0&0 \end{pmatrix}\,  .
\ee
One may easily check with (\ref{assocF}) that the associated ${\cal F}$ is indeed 
of the required form (\ref{finalmonodromy}). 
Since this transformation is integer-valued it 
is indeed compatible with the torus identifications. 
This suffices to  show that in DFT the space is globally well-defined. 
 It is important to note that
the above matrix has determinant $-1$. Thus, this coordinate transformation is orientation-reversing, 
and so realizing this fibered torus as a globally well-defined (doubled) space  forces us 
to view it as some kind of  generalized higher-dimensional M\"obius strip. 
There are actually various different forms of valid coordinate transformations leading to the 
required field transformation, but we have checked that they have all determinant $-1$.  
Therefore, $\det\frac{\partial X'}{\partial X}=-1$ and hence this generalized coordinate transformation 
is \textit{not} connected to the identity, i.e., it is a truly `large' transformation, even though 
 the associated 
${\cal F}(A)$ in (\ref{finalmonodromy})  is actually in the identity component of $O(D,D)$. 
A cautionary remark is in order. 
We are  missing  a characterization of the allowed 
large generalized coordinate transformations, accounting properly
for the strong constraint.\footnote{The definition of the strong constraint
has only been given for coordinate transformations connected to the identity, in which case one writes $X' = X - \xi(X)$ and tests that
$\xi(X)$ satisfies the strong constraint.  The above transformation does
 not satisfy this constraint (as hinted in \cite{Berman:2014jba}):
  $\partial_M \xi^1 \partial^M \xi_1 \not=0$. 
Since the transformation is not connected to the identity, however,  it is unclear if the  failure of the naive strong constraint is
problematic.} 
We leave this subtle problem for future investigations.

\medskip

Let us finally mention that one may construct further non-geometric backgrounds that are even more exotic and that do not fit into the 
strongly constrained DFT known so far. For instance, recently a background has been constructed that contains a genuine 
$R$-flux and for which one accordingly expects only a non-local description, i.e., one which is not covered by T-folds. 
Specifically, this
background can be constructed as an asymmetric $\mathbb{Z}_4\times\mathbb{Z}_2$ orbifold \cite{Condeescu:2013yma}.
One feature of these backgrounds is that the expressions for the K\"ahler and complex structure moduli are inherently non-local
in the sense that they depend simultaneously on the coordinate $z$ and its dual $\tilde{z}$. As such these 
field configurations even violate the weak constraint, which requires $\partial_z\partial_{\tilde{z}}$ to annihilate all 
fields. The weak constraint is equivalent to the level-matching constraint for the massless string fields on the torus, 
but once we consider different backgrounds or include additional fields/states (say, in form of extra gauge fields) 
it is conceivable that there may be constraints that are compatible with these novel backgrounds. 
For the moment this remains as a very non-trivial open problem.

\section{Comments on the strong constraint}\label{STRONGER}
In this section we discuss the role of the strong constraint 
and the issues related to attempts 
to relax it. There are various reasons to believe that the strong constraint can and should be relaxed,
ranging from string theory on a torus background to massive and gauged deformations of supergravity.
We discuss the example of massive type IIA DFT, where 
 the consistency of a weaker 
constraint with the gauge symmetries is simple to understand.

There have been a series of papers   
discussing the construction of gauged supergravity in lower dimensions 
by means of a generalized Scherk-Schwarz 
compactification of 
DFT~\cite{Aldazabal:2011nj,Geissbuhler:2011mx,Geissbuhler:2013uka}.\footnote{We are grateful
to D.~Marques for a discussion and correspondence on this subject.} 
Specifically, the Scherk-Schwarz ansatz allows for a dependence on 
the (doubled) internal coordinates and seems to lead naturally to gauged supergravity 
formulated with the embedding tensor technique~\cite{Nicolai:2000sc,deWit:2002vt}. 
It turns out that in order to obtain the 
most general lower-dimensional gauged supergravity theories
a relaxation of the  
constraints, strong and weak,  is required as one needs a certain restricted dependence on both the internal coordinates and their duals.   
In addition, the DFT action must be supplemented
by a set of terms that would vanish if the 
constraints are enforced.
The resulting lower-dimensional gauged supergravity is consistent, 
in particular gauge invariant, and in this sense the supplemented
DFT action with Scherk-Schwarz ansatz   
is consistent with a relaxed constraint.  
In fact,  \cite{Geissbuhler:2013uka} states 
that once the constraints required by closure of the gauge 
algebra are imposed, 
gauge invariance of the action can be established modulo the very same constraints. 
It would be useful to 
characterize explicitly and understand the scope of the proposed
set of weaker constraints, 
in particular, because the weak constraint is a constraint in string theory. 
We hope that the simpler example of massive type IIA may be a 
guide in order to arrive at a conceptual 
 understanding of generalized Scherk-Schwarz compactifications in DFT.

\subsection{Implications of the strong constraint}   
The constraint (\ref{STRONG}) is interpreted in the strong sense that $\partial^M\partial_M$ annihilates all 
fields and gauge parameters, but also all products, so that for generic 
fields or parameters  
$A$ and $B$ we require 
 \be\label{strongeq}
  \partial^M\partial_M A \ = \ \partial^M\partial_M B \ = \ 0\;, \qquad \partial^MA\,\partial_M B \ = \ 0\;. 
 \ee   
The first group of conditions is referred to as the weak form of the constraint.
The first, together with the second, is referred to as the strong constraint. 

As has been shown in \cite{Hohm:2010jy}, the strong constraint  
 implies that, effectively,  all fields depend only on half 
of the coordinates. They may depend only on $x^i$ or only on $\tilde{x}_i$ or any combination obtained 
by an arbitrary $O(D,D)$ transformation. The subspaces corresponding to these restricted 
coordinates are also called totally null ---  
as all tangent vectors are null with respect to the $O(D,D)$ invariant metric in (\ref{STRONG}). 
Therefore, we can state the strong constraint equivalently as follows: \\[1ex]
\textit{Strong constraint: 
DFT  fields only depend on the coordinates of a totally null subspace.}\\[1ex]
 The conditions (\ref{strongeq}) are then a direct consequence of the strong constraint in this formulation.
 It is noteworthy that in explicit computations only the form  (\ref{strongeq}) is ever required.  The fact that (\ref{strongeq}) implies the above statement of the strong
 constraint is more nontrivial and we discuss the proof now. 
 For this purpose, we first recall Witt's theorem
 on vector spaces:
 
 \noindent
 {\em  Witt's Theorem:  }  Let $V$ be a finite dimensional vector space with a non-degenerate bilinear form.  Any isometry between two subspaces of $V$ can be extended to an isometry of $V$. 
 
 \medskip
 For our setup the vector space will be the $2D$-dimensional
 space $\mathbb{R}^{2D}$ of generalized momenta 
\be
 P^M  \ = \  \begin{pmatrix} p_i \\ w^i \end{pmatrix}\;, 
\ee  
and the non-degenerate bilinear form
 will be the $O(D,D)$ metric $\eta_{MN}$.  Isometries of $V$ are $O(D,D)$
 transformations.    Consider a field $A(x, \tilde x)$ with a single
 Fourier mode
 \be
 A (x, \tilde x) \ = \  A_P\, e^{ip_i x^i + i 
  w^i \tilde x_i} \ = \ A_P  \,  e^{i P_K X^K} \,.
 \ee
 The constraint $\partial^M \partial_M =0$ gives  
 \be
 \eta^{MN} P_M P_N \ = \ 0  \;\; \to \;\; P \cdot P \ = \ 0  \qquad \, P \hbox{ is null}\,. 
 \ee
 If we have two fields $A = A_P e^{iP_K X^K}$ and $A' = A_{P'} e^{iP'_K X^K}$, both $P$ and $P'$ must be null, but
 the strong constraint $\partial^M A \, \partial_M A' =0$ implies that, in addition, these two momenta must be orthogonal
 \be
 P \cdot P' \ =\ 0 \,. 
 \ee
 It is thus the case that all momenta appearing in fields or gauge parameters
span an $N$-dimensional
 isotropic subspace $\mathbb{S}^N$ of the momentum space 
$\mathbb{R}^{2D}$, namely,
a subspace of null vectors.  Indeed,  since all momenta appearing on fields or gauge
parameters must be null and any chosen pair must be orthogonal, any linear superposition of these momenta is null.  Since isotropic subspaces of $\mathbb{R}^{2D}$ with metric $\eta$ cannot have dimensionality larger than $D$, we must have $N \leq D$.  
 
Now consider the maximal isotropic subspace $\mathbb{E}^D$ of  $\mathbb{R}^{2D}$ described
by a basis of $D$ vectors
without winding
\be
E_1 =\begin{pmatrix} e_1 \\ 0 \end{pmatrix} \,, \ \ E_2 =\begin{pmatrix} e_2 \\ 0 \end{pmatrix} \,, \ \ E_3 =\begin{pmatrix} e_3 \\ 0 \end{pmatrix}  \,, \  \ldots\,,\  \  E_D = \begin{pmatrix} e_D \\ 0 \end{pmatrix} \,. 
\ee 
We argued above that the momenta that appear on the fields that satisfy the strong constraint span the isotropic space $\mathbb{S}^N $. Let $V_i$, with $i= 1, 2, \ldots N$ denote a basis
for $\mathbb{S}^N$.  Consider now the linear map $m_N$ defined by the map of basis vectors
\be
m_N : \ \ V_i \to  E_i  \,, \ \ \   i = 1 ,2 , \ldots , N\,.
\ee  
This map is clearly an isometry between $\mathbb{S}^N$ and $m_N(\mathbb{S}^N)$, both of which are subspaces of $\mathbb{R}^{2D}$.
It follows by Witt's theorem that $m_N$ can be extended
to an $O(D,D)$ transformation of $\mathbb{R}^{2D}$.  Therefore there is an $O(D,D)$ transformation that
maps all relevant momenta to a subspace of vectors without winding.   Fields without
winding are fields that do not depend on $\tilde x$.  This shows that there is 
an $O(D,D)$ transformation to a coordinate 
frame in which fields do not depend on the tilde coordinates.

 \subsection{Can the strong constraint be relaxed?}  
 
 Let us now address the question: Can the strong constraint be relaxed? Asked in this generality the 
 answer is undoubtedly yes. In fact, in closed string field theory on a torus background only the 
 level-matching constraint is required, which is a weaker form of (\ref{strongeq}). For the massless 
 fields (for which the number operators $N$ and $\bar N$ both have eigenvalue
 equal to one) it reads 
  \be
   L_0-\bar{L}_0 \ = \ -p_i w^i \ = \ 0\;, 
  \ee
 where $p_i$ and $w^i$ are the momentum and winding modes, respectively. 
 Translating to 
 coordinate space this constraint implies $\tilde{\partial}^i\partial_i=0$ and thus $\partial^M\partial_M=0$ 
 when acting on the massless fields and the associated gauge parameters. 
 In general, $L_0-\bar{L}_0  = N- \bar N -p_i w^i =0$, 
 so massive fields 
 have integrally quantized values of $\partial^M\partial_M$.  Closed string field theory is in fact a fully consistent 
 weakly constrained DFT, and so in the full string theory the doubled coordinates are undoubtedly 
 physical and real. Of course, the full string theory is quite intricate and it is
 therefore of interest to focus just on the massless sector.  
 For this sector, we can define the strong constraint by the requirement that the
 operator $\partial^M\partial_M$ that annihilates fields and gauge parameters, also
 annihilate all products of fields and/or gauge parameters.\footnote{It is not 
 clear if there is a useful definition of strongly constrained string field theory.} 
 Therefore, the more interesting question is this: Can the strong constraint be relaxed for the massless string fields only
 and/or to a finite order in $\alpha'$? In the following we address the issues one encounters in relation to 
 this question. 
 
 There are two main obstacles one encounters when trying to relax the strong constraint to the weak   
 constraint.\footnote{One can try to relax both, as is the case in the works of
 \cite{Aldazabal:2011nj,Geissbuhler:2011mx,Geissbuhler:2013uka}.}
 First, one has to 
 find an action and gauge transformations so that invariance of the action and closure of the gauge algebra 
 require only the use of $\partial^M\partial_MA=0$, not of $\partial^MA\,\partial_MB=0$.  
This is a very non-trivial problem, as the latter condition is heavily used in most DFT computations.   
 But the second obstacle is even more severe:  
one must make the symmetry 
 variations consistent with the weak constraint. 
More precisely, if we impose $\tilde{\partial}^i\partial_i\Phi=0$ for a generic DFT field $\Phi$ or a gauge parameter,  
consistency 
requires that the gauge variations, which read schematically 
 \be
 \delta_{\xi}\,\Phi \ = \ \xi\cdot \Phi\;, 
 \ee
 should respect the constraint. Since we no longer have 
the strong constraint, the product of parameter and field in general does not satisfy the constraint, so that 
  \be\label{constrproblem}
   \tilde{\partial}^i\partial_i(\delta_{\xi}\Phi) \ \neq \ 0\;. 
  \ee
Thus the gauge variation is not compatible with the weak constraint.   
In order to remedy this we have to project out those Fourier modes that violate the 
constraint $p\cdot w=0$. Denoting this projection by $[\,\,]$ we may try a new ansatz for the 
gauge transformations,
 \be
   \delta_{\xi}\,\Phi \ = \ \big[\,\xi\cdot \Phi\,\big]\;.
 \ee
Since we are projecting out unwanted Fourier modes, this modification of the gauge transformations 
introduces a non-locality into the theory. This projection and the associated non-locality are, however, 
present in closed string field theory as well, as this is the way it is consistent with the weaker level-matching 
constraint. 
After introducing the projectors,  
gauge invariance of the action and closure of the gauge algebra become a highly non-trivial issue and can only be established 
if the projector satisfies a sufficient number of algebraic identities, 
perhaps exhibiting a structure similar to the $L^{\infty}$-algebras governing closed string field 
theory \cite{Zwiebach:1992ie}.\footnote{In collaboration with C.~Hull two of us (OH and BZ) 
have obtained partial results along these lines, but they are as yet inconclusive.}

 The first obstacle of formulating 
gauge transformations whose gauge algebra  closes and that leave an action invariant modulo only $\partial^M\partial_MA=0$ 
has actually been solved in one particular case, the DFT for the
Ramond-Ramond (RR) sector
of type II. 
 The second obstacle can be overcome by a partial weakening of 
 the strong constraint.  We discuss this now.

\subsection{Massive type II: minor relaxation of the strong constraint}
We start by recalling the basics of type II DFT as constructed in \cite{Hohm:2011zr,Hohm:2011cp}. 
The RR fields are encoded in a Majorana-Weyl spinor of the two-fold covering group Spin$(10,10)$ of $SO(10,10)$, while 
the generalized metric is uplifted to an element $\fancys$ of Spin$(10,10)$, satisfying $\fancys  =  \fancys^{\dagger}$. 
The Clifford algebra reads 
 \be\label{Clifford}
  \big\{ \Gamma^{M},\Gamma^{N}\big\} \ = \ 2\eta^{MN}\,{\bf 1}\;, 
 \ee
so that due to the off-diagonal form of the $O(D,D)$ metric the gamma matrices 
can be identified with fermionic lowering and raising operators  $\psi_{i}$ and $ \psi^{i}$, respectively,  
  \be \label{defgamma}
   \psi_{i} \ \equiv \ \frac{1}{\sqrt{2}}\,\Gamma_{i}\;, \quad
   \psi^{i} \ \equiv \ \frac{1}{\sqrt{2}}\,\Gamma^{i}\;,
 \ee
with $(\psi_i)^{\dagger}=\psi^i$. We can thus introduce a Clifford vacuum $\ket{0}$, satisfying $\psi_i\ket{0}=0$ for all $i$,
and define the spinors by acting with the raising operator. The RR $p$-form potentials $C^{(p)}$ 
are then encoded in the spinor 
   \be \label{genstate}
  \chi \ = \ \sum_{p=0}^{10}\frac{1}{p!}\,C_{i_1\ldots i_p}\,\psi^{i_1}\ldots\psi^{i_p}\ket{0}\;. 
 \ee
The Dirac operator corresponding to the Clifford algebra (\ref{Clifford}), 
 \be\label{def-dir-op}
  \slashed\partial \ \equiv \  {1\over \sqrt{2}} \, \Gamma^M \partial_M
  \ = \   \psi^i\partial_i+\psi_i\tilde{\partial}^{i} \,,
 \ee
then acts on the spinor (\ref{genstate}) as a natural $O(D,D)$ covariant extension of the 
exterior derivative. In fact, for $\tilde{\partial}^i=0$ it acts on (\ref{genstate}) by increasing the 
form degree by one and taking the totally antisymmetrized derivative, exactly as the 
exterior derivative on forms. Moreover, it generally squares to zero thanks to the first constraint in (\ref{strongeq}), 
\be\label{zerosquare}
  \slashed{\partial}^2 \ = \
  \frac{1}{2}\Gamma^{M}\Gamma^{N}\partial_{M}\partial_{N} \ = \
  \frac{1}{2}\eta^{MN}\partial_{M}\partial_{N} \ = \ 0\,,
 \ee
where we used (\ref{Clifford}). 
Below we will need this relation only in the weak form, when 
$\slashed{\partial}^2$ acts on a single field or parameter.

The complete bosonic type II DFT action  reads   
 \be
\label{totaction}
 S \ = \  \int d^{10}x\, d^{10}\tilde x\, \Bigl(  e^{-2d}\, {\cal R} ({\cal H} , d)  + \frac{1}{4}
 (\slashed{\partial}{\chi})^\dagger \;
 \fancys \; \slashed\partial\chi\,\Bigr) \,,
\ee
and is supplemented by the self-duality constraint  
 \be\label{dualityintro}
 \slashed\partial\chi \ = \ -{\cal K}\,\slashed\partial\chi\,, \quad
 {\cal K} \ \equiv \ C^{-1} \fancys\, ,
\ee
where $C$ denotes the charge conjugation matrix of Spin$(10,10)$. 
This theory is gauge invariant under the generalized diffeomorphisms parametrized by $\xi^M$, 
which act on the spinor $\chi$ as 
 \be\label{delta0chi}
    \delta_\xi \chi \ = \ \xi^M \partial_M  \chi +  {1\over 2}   \partial_M \xi_N \Gamma^M
  \Gamma^N \chi\;, 
 \ee
and under a new abelian gauge symmetry with a Spin$(10,10)$ spinor parameter $\lambda$, 
 \be\label{abelian}
  \delta_{\lambda} \chi \ = \ \slashed{\partial} \lambda\,, 
 \ee
which reduces to the usual $p$-form gauge symmetry when $\tilde{\partial}^i=0$.  
The invariance of $\slashed{\partial}\chi$ under this symmetry (and thus the invariance of the action and self-duality constraint) 
is manifest thanks to  $\slashed{\partial}^2=0$.

As usual, the above action is only $\xi^M$ gauge invariant if we impose the strong constraint, 
but as noted above $\slashed{\partial}^2=0$ and thus $\lambda$ invariance requires only the 
weaker constraint. 
It turns out that we can also reformulate the $\xi^M$ gauge transformations of the RR fields so as 
to allow for a minor relaxation of this constraint. To this end we rewrite $\delta_{\xi}\chi$ as 
  \be
   \delta_{\xi}\chi \ = \ 
    \xi^{M}\partial_{M}\chi-\frac{1}{2}\Gamma^{M}\Gamma^{N}\xi_{N}\partial_{M}\chi
  +\frac{1}{2}\Gamma^{M}\partial_{M}\left(\xi_{N}\Gamma^{N}\chi\right)\;. 
 \ee
 The last term in here takes the form of a field-dependent abelian  
gauge transformation with  parameter $\lambda=\frac{1}{\sqrt{2}}\xi^N\Gamma_N\chi$, and so upon 
a field-dependent parameter redefinition this term can be eliminated. 
Using then the Clifford algebra in the remaining term above we arrive at 
\be\label{finalgauge}
  \delta_{\xi}\chi \ = \ \slashed{\xi}\,\slashed{\partial}\chi\;,
 \ee
where we set  $\slashed{\xi}  =  \frac{1}{\sqrt{2}}\Gamma^{M}\xi_{M}$. 
The claim is now that closure of the gauge algebra on $\chi$ and $\xi^M$ and  gauge 
invariance of the RR action require only the weaker constraint. More precisely, these 
computations require the conditions 
 \be\label{weakconstr}
  \partial^{M}\partial_{M}A\ = \ 0\;, \qquad
  A \ = \ \big\{\chi,\lambda,\xi^{M}\big\}\;, 
 \ee 
but no condition of the form $\partial^MA\,\partial_MB=0$. 

We thus have shown that for the RR sector the first of the obstacles discussed in the previous subsection 
has been solved. However, as stressed above in regard to the second obstacle, 
this is not yet sufficient to claim to have a weakly constrained theory. In fact, without the strong constraint 
the gauge transformations (\ref{finalgauge}) are not consistent with the weak constraint, for the same 
reasons as in (\ref{constrproblem}), but we can still relax the strong constraint somewhat using the 
observation that in contrast to (\ref{delta0chi}) in  (\ref{finalgauge}) the field appears only under a 
derivative. Therefore, if we take the RR fields   
to be a sum of terms that depend arbitrarily on $x^i$,  we can also include terms
that depend linearly on $\tilde x_i$ but have no $x^i$ dependence.  
The gauge variation will then be $\tilde{x}$ independent and thus 
compatible with (\ref{weakconstr}).  
We have therefore arrived at the following weakened version of the 
strong constraint: \\[1ex]
\textit{Weakened strong constraint: 
The RR fields of DFT can only depend on the coordinates of a totally null subspace and at most linearly on the 
coordinates orthogonal to this space.}

It turns out that thanks to this relaxation the type II DFT (\ref{totaction}) now also encodes 
the massive type IIA theory due to Romans \cite{Romans:1985tz}. To see this we have to make an ansatz for the 
RR one-form that depends linearly on one of the $\tilde{x}_i$, say $\tilde{x}_1$, 
 \begin{equation}\label{ansatz} 
  C^{(1)}(x,\tilde{x}) \ = \  C_{i}(x)dx^{i}+m\tilde{x}_{1} dx^{1}\;, 
 \end{equation}
where $m$ is a mass parameter.  
Inserting this into the field strength $F$ we obtain 
  \begin{equation}
  F \ = \  
  \slashed{\partial}\chi \ = \  (\psi^i\partial_i+\psi_i\tilde{\partial}^i)\chi
 \ = \ F_{m=0}+\psi_i\tilde{\partial}^i(m\tilde{x}_1)\psi^1\ket{0}\;.
 \end{equation}
Note that 
the $\slashed{\partial}$ operator acts non-trivially on the 
$m$-dependent part. 
In this last term  the two operators $\psi^i$ and $\psi_i$ annihilate each 
other, leading to a non-vanishing contribution to the `zero-form' field strength:
  \begin{equation}
   F^{(0)} \ = \  m \;. 
 \end{equation}
Thus, 
the $\tilde{x}$-dependent part acts as a `$(-1)$-form' in the sense that acting 
with the (generalized) exterior derivative $\slashed{\partial}$ we obtain a zero-form. 
It has been noticed before in the literature that if one formally introduces $(-1)$-forms the
formulation of massive supergravity simplifies \cite{Lavrinenko:1999xi}.
Here we obtained a natural geometric interpretation of these exotic objects:
 $(-1)$-forms are 1-forms depending on the dual coordinates. 
 Insertion of the ansatz (\ref{ansatz}) into type II DFT indeed precisely  reproduces 
 the massive type IIA theory \cite{Hohm:2011cp}. 

We have thus 
seen that for the RR subsector of DFT the strong constraint can be 
relaxed somewhat so as to allow for a simultaneous dependence on coordinates and their 
duals, provided one of them enters only linearly. The 
resulting conventional spacetime 
theory then corresponds to a massive deformation.

\section{Doubled $\alpha'$ geometry}

As the generalized metric form of double field theory was developed it became
clear that one could try to include higher-derivative terms in the action while
preserving the continuous $O(D,D)$ symmetry of the theory as well as the gauge
symmetries~\cite{Hohm:2010pp}.   
It was soon realized,  
however, that the expected higher derivative
corrections, in particular, the Riemann-squared terms were difficult to include while
preserving the symmetries of the theory  \cite{Hohm:2011si}.

This difficulty makes it instructive to discuss why we expect that the continuous
$O(D,D)$ symmetry of the two-derivative DFT survives the inclusion of $\alpha'$ corrections from string theory.  For this we will review an argument by Sen\cite{Sen:1991zi}.   We then discuss and  explain a few of the key results of
the recent construction of $\alpha'$ corrected DFT presented in~\cite{Hohm:2013jaa}.

\subsection{$O(d,d)$  survives $\alpha'$ corrections}
We now consider the argument of~\cite{Sen:1991zi}, which explains why a string theory with $d$ space-like coordinates described by 
 $d$ free scalar fields leads to a reduced string theory with $O(d,d)$ continuous 
 symmetry.  This is true including $\alpha'$ corrections.   This result makes it plausible
 that a DFT with global $O(D,D)$ must exist upon inclusion of $\alpha'$ corrections.

For the standard matter CFT of free bosons the holomorphic and antiholomorphic sectors decouple.  All correlators are invariant under simultaneous and independent
rotations of the $\partial X^i$ and $\bar \partial X^i$ currents, with $i= 1 ,2, \ldots , d$.  Sen's argument is couched in the language of string field theory and effective
actions.  It begins by
stating that the string field encoding the fluctuations of the (internal) metric and $b$-field takes the form
\be
(h_{ij} + b_{ij})\, \alpha^i_{-1}  \bar\alpha_{-1}^j  c_1 \bar c_1 |0\rangle \;.
\ee 
As a consequence of the correlators' invariance, the string field theory will have the exact symmetry under
the $O(d) \times O(d)$ action 
\be
\label{xcv}
h + b \ \to \ \tilde h + \tilde b  \ = \ S ( h + b) R^T\,, ~~\ \ \ \  S, R \in O(d)  \,. 
\ee
Consider now the effective field theory of the fields $G_{ij}, B_{ij}$ as known
with two derivatives.  This theory uses a generalized metric ${\cal H}$ as in (\ref{firstH}) with
$(g,b)$ replaced by  $(G, B)$, and has a symmetry
\be
\label{calxh}
{\cal H}\  \to \  \tilde {\cal H} \ = \  \Omega \,  {\cal H}\  \Omega^t\,,
\ee
where $\Omega\in O(d,d)$.  There is an $O(d) \times O(d)$ subgroup 
described by the matrices
\be
\label{omex}
\Omega \ = \ {1\over 2} \begin{pmatrix}   S+R  &  R-S \\ R- S  & S+ R \end{pmatrix} \;. 
\ee
This subgroup leaves invariant the flat background $G_{ij} = \delta_{ij}, \, B_{ij} =0$.
Since we view the string field as fluctuations of $G, B$ around the flat background, we have
\be
\begin{split}
G_{ij} \ = \ & \ \delta_{ij} + h_{ij}  + \ldots \;, \\
B_{ij} \ = \  &  \  b_{ij} + \ldots\;, 
\end{split}
\ee
where the dots indicate terms quadratic and higher order in the fluctuations. It is now possible to verify that the symmetry (\ref{calxh}), (\ref{omex}), through the above
expansion turns into the symmetry (\ref{xcv}).  Since this $O(d)\times O(d)$ of
the string field theory is exact and holds to all orders of $\alpha'$ the symmetry should
exist in the low-energy theory described with fields $(G, B)$.  This subgroup contains 
$2\cdot  {1\over 2}\cdot  d (d-1) = d (d-1)$ generators.  

We then note that any coordinate invariant theory with gravity and
a dilaton (such as the reduced theory we
are considering) has a $GL(d)$ symmetry arising from a $GL(d)$ transformation
of the $d$ coordinates: $x^i \to A^i{}_j \, x^j$.  This symmetry should exist
to all orders in $\alpha'$.  At the level of the $G, B$ fields we
have $G \to A G A^T$ and $B \to A B A^T$, while the dilaton is shifted as
$\phi \to \phi + \ln (\hbox{det} A)$.   In terms of $O(d,d)$ these transformations
arise from 
\be
\label{omep}
\Omega \ = \ 
\begin{pmatrix}   A  &  0 \\ 0  & (A^t)^{-1}  \end{pmatrix} \,.
\ee
 Not all of the $GL(d)$ generators are new: the diagonal subgroup $O(d)$
 of $O(d) \times O(d)$
  with $R = S$  is also a subgroup of $GL(d)$.  Thus only $d^2 - {1\over 2} d(d-1) = {1\over 2} d (d+1)$ generators are new.  
  
  Finally, the low-energy theory theory must have, even with $\alpha'$ corrections,
  the Kalb-Ramond symmetries
  $\delta B_{ij} = \partial_i \epsilon_j - \partial_j \epsilon_i $ of the two-derivative theory.  Taking $ \epsilon_i  = f_{ij} x^j$ with constant $f_{ij}$'s  we get 
  $\delta B_{ij} = f_{ji} - f_{ij} \equiv C_{ij}$, with $C$ an antisymmetric constant
  matrix. This 
  transformation is an $O(d,d)$ transformation with 
  \be
  \Omega \ = \ \begin{pmatrix}   1  &  0 \\ C  & 1  \end{pmatrix} \,.
  \ee
 These so-called $B$-shifts amount to ${1\over 2} d(d-1) $ generators.  With
 $O(d) \times O(d)$, the additional generators from $GL(d)$ and the $B$ shifts
 we have identified all the generators of $O(d,d)$.  Indeed the count works:
 \be
 d(d-1)  \ + \ {1\over 2} d (d+1)  \ + \ {1\over 2} d (d-1)  \ = \ 2d^2 - d \ = \ {1\over 2}  (2d) (2d-1) \;, 
 \ee
 which is the number of generators of $O(d,d)$.   This concludes the argument that the
 reduced theory of massless fields must have a global, continuous $O(d,d)$ symmetry even as $\alpha'$ corrections are included.

  \subsection{$\alpha'$-geometry}
  
The absence of a duality covariant generalized Riemann tensor that is fully determined
in terms of the physical fields of DFT, 
see \cite{Hohm:2011si}, made it clear that some deformation of the
structures of the theory was needed in order to include the Riemann-squared 
terms that arise in the $\alpha'$-expansion of the effective theory for
the massless sector of closed strings.  It was generally believed that the duality
transformations of the fields would be modified while the gauge transformations, 
which include diffeomorphism and b-field transformations, would not be changed.
Nevertheless, it was anticipated from string field theory that the opposite would be true: duality would remain
manifest (thus un-corrected) while gauge symmetries would receive some kind of corrections.

Indeed, the formulation of~\cite{Hohm:2013jaa} 
shows that the gauge structure
of the theory is changed.  Recall that gauge parameters $\Xi^M (X)$ contain 
components  $(\tilde \xi_i, \xi^i)$ that depend both on ($x, \tilde x)$.
When we restrict ourselves to fields and gauge parameters that do not depend on
$\tilde x$,  $\tilde \xi_i$ becomes a one-form and $\xi^i$ a vector.   

In the two-derivative theory (no $\alpha'$ corrections)
we had the following key structures
\be
\begin{split}
 \langle \Xi_1|\Xi_2\rangle  \ = \ \ & \xi_1^M \xi_2^N \eta_{MN} \,, \\ 
[ \Xi_1 , \Xi_2 ]_{{}_C}^M  \ = \ \ & \xi_{[1}^N \partial_N \xi_{2]}^M \, - \, {1\over 2}  \,
\xi_1^K \, \overleftrightarrow{\partial}^{\hskip -1pt M}  \xi_{2K}\,, 
\\
\widehat{\cal L} _\Xi  V^M  \ = \ \  &  \xi^{P}\partial_{P}V^{M}
  +(\partial^{M}\xi_{P}\,    -\partial_{P}\xi^{M})  V^{P}
 \,. 
\end{split}
\ee
(In this section only we use the (anti)symmetrization convention $[ab]=ab-ba$,  
and $A \overleftrightarrow{\partial} B  \equiv A \partial B - (\partial B) A$.)
The first structure is the inner product, formed simply using the metric $\eta$.  The
inner product is a generalized scalar formed from two vectors.  The second structure is
the C-bracket of two gauge parameters (vectors), giving a third.  When restricted to parameters that do not depend on $\tilde x$, it reduces to the Courant bracket of generalized geometry.  The third line defines 
the generalized Lie derivative of a generalized vector $V$ along $\Xi$.  The commutator of two such Lie derivatives, with two different gauge parameters, is a Lie derivative with respect to a parameter formed by 
taking the C-bracket of the two original  parameters. 
 
When we include $\alpha'$ corrections the above structures are modified.  Interestingly, 
the modifications that were obtained are exact:  they do not represent just first-order corrections to be supplemented by further, or infinitely many,  terms  to be determined. They are complete and self-consistent corrections.  We do not write them with explicit factors of $\alpha'$ but rather the corrections are recognized by the higher number of derivatives; a factor of $\alpha'$ is associated with two derivatives.  
 The new structures, denoted the same way as the ones above except for the
 Lie derivative, are given by
 \be
 \label{corr-structures}
\begin{split}
 \langle \Xi_1|\Xi_2\rangle  \ = \ &\, \xi_1^M \xi_2^N \eta_{MN} - (\partial_N \xi_1^M)(\partial_M \xi_2^N)\,, \\ 
[ \Xi_1 , \Xi_2 ]_{{}_C}^M  \ = \ &\, \xi_{[1}^N \partial_N \xi_{2]}^M \, - \, {1\over 2}  \,
\xi_1^K \, \overleftrightarrow{\partial}^{\hskip -1pt M}  \xi_{2K}\, + \, {1\over 2}  
\, (\partial_K \xi_1^L) \overleftrightarrow{\partial}^{\hskip -1pt M} (\partial_L \xi_2^K)\,,
\\
{\bf L}_\Xi  V^M  \ = \  &\,  \xi^{P}\partial_{P}V^{M}
  +(\partial^{M}\xi_{P}\,    -\partial_{P}\xi^{M})  V^{P}
  -\,   (\p^M \partial_K \xi^L  ) \partial_L V^K \,. 
\end{split}
\ee
Each structure has one additional term: the inner product has a two derivative
term and both the bracket and the generalized Lie derivative now have a term
with three derivatives.  The commutator of two corrected Lie derivatives gives, exactly, a corrected Lie derivative with parameter given by the corrected C bracket.
Moreover the Jacobiator of the corrected C bracket is still a trivial gauge parameter,
as it was for the two-derivative theory and as needed for consistency.
The correction to the generalized Lie derivative implies that the transformation
law  $V' = {\cal F} V$ with ${\cal F}$ given by (\ref{GenTRAF}) requires modification
to order $\alpha'$.  We will not attempt here to find such modification.

Interestingly, the  corrections in (\ref{corr-structures}) do not vanish when we
restrict to fields and parameters that do not depend on $\tilde x$.  Therefore
the above corrections imply corrections to the familiar structures of generalized
geometry.  Consider, for example, the inner product,  which in the absence of 
$\tilde x$ dependence becomes
\be
\langle \Xi_1 | \Xi_2\rangle  \ = \ \xi_1^i \tilde\xi_{2i} + \xi_2^i \tilde\xi_{1i}  \ - \  \p_i \xi_1^j  \, \p_j \xi_2^i \,. 
\ee
The last term is 
the higher-derivative correction.  
For the C bracket the vector part is not corrected,
but the one-form part is
\be
\begin{split}
\left( [ \Xi_1 , \Xi_2 ]_{{}_C}\right)^i  \ =\ &  \  \xi_{[1}^k \p_k \xi_{2]}^i    \,, \\[0.5ex]
\left( [ \Xi_1 , \Xi_2 ]_{{}_C}\right)_i  \ = \ &  \ \xi_{[1}^k \p_k \tilde\xi_{2]i} \, -  
{\textstyle {1\over 2}} 
\bigl( \xi_1^k \overleftrightarrow{\partial_i} \tilde \xi_{2k} + \tilde \xi_{1k} \overleftrightarrow{\partial_i} \xi_2^k \bigr) \ + \  {\textstyle {1\over 2}} 
\, (\partial_k \xi_1^\ell)\overleftrightarrow{\partial_i} (\partial_\ell \xi_2^k)  \,.
\end{split}
\ee
The correction term is the last term on the right-hand side of the second line. 
Finally, for the Lie derivatives we have 
\be
\label{corrliegen}
\begin{split}
({\bf L}_\Xi  V)^i   \ = \ & \ \xi^k \p_k V^i - V^k \p_k \xi^i   \\
({\bf L}_\Xi  V)_i  \ = \ & \ \xi^k \p_k V_i\, + \, \p_i \xi^p \,V_p  + (\p_i \tilde \xi_p - \p_p \tilde \xi_i) V^p \, - \, \p_i \p_k \xi^p \p_p V^k \,.
\end{split}
\ee
The correction is the last term on the second line and it only affects the Lie derivative
of the one-form. Note, however, that this correction involves the vector $V^k$.  Thus
at order $\alpha'$ the one-form and the vector mix under generalized
diffeomorphisms!  

Given the gauge structure above, reference~\cite{Hohm:2013jaa} 
looked for
fields and gauge transformations. The key technical tool is a chiral CFT in
the doubled space with a novel propagator and simplified OPE's, as the
result of the strong constraint.  The CFT has currents $Z^M(z) \equiv {X'}^M (z)$
and the dynamical fields are introduced through the weight-two operators
\be
\begin{split}
{\cal S} \ \equiv \ & \  \ {1\over 2}   (Z^2 - \phi'')\;,  \\
{\cal T}  \ \equiv \ & \  {\textstyle {1\over 2}}
 {\cal M} ^{MN} Z_M Z_N  -  {\textstyle {1\over 2}} (\widehat {\cal M}^M Z_M)' \,.
\end{split}
\ee
Here we introduced the dilaton $\phi=-2d$, in order to comply with the conventions of \cite{Hohm:2013jaa}, 
which enters in ${\cal S}$, while  
the {\em double metric} ${\cal M}$ enters
in ${\cal T}$.  The double metric is related but is not equal to the generalized metric 
${\cal H}^{MN}$. While off-shell ${\cal H}$  squares to one, the former is unconstrained. The second term in ${\cal T}$, needed for consistency with gauge
transformations, contains an auxiliary field $\widehat {\cal M}$.  This field  is  determined in terms of 
the double metric and the dilaton by the condition 
$\hbox{div}\, {\cal T} =0$, where the divergence of an operator
 is defined through its OPE with ${\cal S}$.  Thus ${\cal T}$ is a divergenceless operator.

Gauge transformations 
of any operator ${\cal O}$ are defined by the
commutator
$\delta_\Xi {\cal O}  =  \bigl[  \int \Xi \,, {\cal O} \bigr]$, and
 are readily evaluated with the use of operator products.   This was used in
  \cite{Hohm:2013jaa} 
  to find the gauge transformations of ${\cal M}$ and $\phi$
\be
\label{gtacfin}
\begin{split}
\delta_\Xi  {\cal M}^{MN} \ = &\ \,  \xi^P \partial_P  {\cal M}^{MN} + ( \partial^{M}\xi_{P} - \partial_{P}\xi^{M}) {\cal M}^{PN}  
 + ( \partial^{N}\xi_{P} - \partial_{P}\xi^{N}) {\cal M}^{NP}  
\\[0.8ex]
	& - {\textstyle {1\over 2}} \, \bigl[ \, \partial^M \hskip-2pt {\cal M}^{PQ}\, \partial_P (\partial_Q \xi^N -\partial^N \xi_Q )  + 2\,\partial_Q  {\cal M}^{KM}\, \partial^{N} \partial_K \xi^Q
	+ (M\leftrightarrow N)  \bigr] 
	\ \   \\[0.8ex]
	&  -\,     {\textstyle {1\over 4}} \,  \partial_K \partial^{(M} \hskip-2pt  {\cal M}^{PQ}\,  \partial^{N)} 
	\hskip-1pt \partial_P \partial_Q \xi^K \,, \\[0.8ex]
\delta_\Xi \phi   \ = &\ \,	 \xi\cdot \p\phi  + \p\cdot \xi \,. 
\end{split}
\ee
The gauge transformation of the double metric ${\cal M}^{MN}$ receives $\alpha'$
and ${\alpha'}^2$ corrections (second and third lines, respectively) since we recognize
that the first line is simply the familiar generalized Lie derivative.  The gauge transformation of the dilaton is unchanged.  

For the dynamics, the key idea is that the equations of motion are  the condition that ${\cal S}$ and ${\cal T}$ satisfy the Virasoro algebra.  A gauge invariant action that implements this idea can be written and it takes the cubic form 
\be S = \int e^\phi \bigl( \,  \langle {\cal T} |{\cal S} \rangle \,  - \, 
{\textstyle {1\over 6}}\langle {\cal T}  |  {\cal T} \star {\cal T} \rangle  \bigr) . \ee
The definition of the various ingredients required 
here were given in~\cite{Hohm:2013jaa}. 
For example, $\langle \cdot  | \cdot \rangle$ above is a
scalar inner-product between weight-two tensors and the  $\star$-product of
two weight-two tensors  gives a divergenceless weight-two tensor as the answer. 
As it turns out, in terms of the double metric, the equation of motion
takes the form ${\cal M}^2  =  1 + 2 {\cal V}$, 
where ${\cal V}$ is  quadratic in ${\cal M}$ and contains from two up to six derivatives.
 While the generalized metric squares to the identity, the double metric ${\cal M}$ squares to the identity up to 
 higher derivatives terms.  The replacement of ${\cal H}$ for the unconstrained
 ${\cal M}$  is a very significant departure
 from the two-derivative theory,  forced by $\alpha'$ corrections.

\section{Conclusions and Outlook} 

We have reviewed DFT with a focus on the 
geometrical aspects such as the notion of 
generalized manifold. We tried to convince the reader that the structures 
required by DFT inevitably 
require a generalization of the manifold structure. 
As the notion of a manifold is deeply ingrained in 
our intuition, this may appear to be a rather radical and speculative 
step, but we should stress that such a conclusion follows quite conservatively from what we 
know about string theory and, more specifically, string field theory. 
More precisely, this conclusion relies on the following inputs:
\begin{itemize}
\item[(i)] Closed string field theory for a torus background
\item[(ii)] Background independence
\item[(iii)] Gauge transformations as some form of coordinate transformations 
\end{itemize}
In fact, string field theory on a torus (input (i)) inevitably leads to doubled coordinates and gauge transformations
 involving doubled derivatives (explicitly known to cubic order). Then requiring a manifestly background independent 
formulation (input (ii)), thereby not requiring anymore a torus background, uniquely leads to the generalized 
Lie derivatives of DFT.  Finally, if one wants to reproduce these generalized Lie derivatives as some sort  
of coordinate transformation $X\rightarrow X'$ (input (iii)), we are forced to adopt a notion of generalized 
coordinate transformations, thus requiring a generalized notion of manifold. Among the above assumptions, 
(iii) is perhaps the one that one could 
possibly imagine to abandon. Perhaps 
in the ultimate 
formulation of string theory the notion of manifold and coordinates 
will disappear altogether, but as long as we  
stick to coordinates, the emergence of a generalized manifold of the type 
discussed here seems inevitable.

The usefulness of DFT and its associated generalized geometric structures 
is evident in  
the striking economy of the corresponding formulations of the low-energy effective actions of string theory. 
The most conservative interpretation of DFT is to treat the winding coordinates as purely formal 
and to think of the doubled derivatives as $\partial_M=(0,\partial_i)$. 
In this case DFT may be viewed as the 
physical implementation of the 
`generalized geometry' program of Hitchin-Gualtieri \cite{Hitchin:2004ut,Gualtieri:2003dx,Gualtieri:2007bq}. 
It is clear, however, that DFT goes beyond generalized geometry  
in at least two aspects: First, certain non-geometric backgrounds, such as those given in eq.~(\ref{eq:nongeofields}) 
and (\ref{nongeometaurho}), 
are not globally well-defined in standard supergravity 
(nor in the generalized geometry rewriting),   
but can be consistently patched together in DFT once we allow for 
generalized coordinate transformations 
that mix $x$ and $\tilde{x}$ coordinates. 
In this way, DFT appears to provide a 
concrete framework for the 
inclusion of `T-folds'.  Second, the $\alpha'$-corrections of string theory, formulated in DFT, lead
to  modifications of the gauge transformations and gauge algebra that are non-trivial even on the 
half-dimensional subspace relevant for generalized geometry~\cite{Hohm:2013jaa}. In particular,
Lie derivatives 
in the $D$-dimensional subspace receive $\alpha'$ corrections as do the 
defining structures of generalized geometry (like the Courant bracket). 

Despite these first glimpses of a radically different geometry underlying string theory, it is clear that 
we are just beginning to understand these new structures and that the presently known DFT and its 
geometry are just first approximations to the full story. 
For instance, the strongly constrained DFT allows 
only for rather 
mild deviations form standard geometry, as in the form of T-folds 
and some non-geometric backgrounds.  Massive IIA gives a clear example of
a possible relaxation of the strong constraint~\cite{Hohm:2011cp}.
With  the recent progress reported in \cite{Geissbuhler:2013uka}  
it is becoming clear that the most general gauged supergravities appearing
in four dimensions may be obtained by a Scherk-Schwarz reduction of 
a suitably extended DFT.  
 The weak and strong constraints are not assumed, instead, a set of 
weaker constraints
arising from the closure of the gauge algebra is imposed. 
The recent reformulation of DFT given in \cite{Hohm:2013nja}
may be helpful  
to understand better this construction in that it puts 
DFT in a form already adapted to general Kaluza-Klein compactifications, 
thus representing the ideal starting point for a comparison 
of the lower-dimensional and higher-dimensional constraints.   
Finally, in regard to the (further generalized)
$\alpha'$-geometry we are just at the beginning, 
and it remains to be seen how the formulation of this 
geometry will enlighten our understanding of string theory in general.

Some further aspects about the covariant 
description of non-geometric fluxes in double field theory
will be presented in \cite{HHLZ}, including
their relation to gauged supergravity and T-folds, as well as a
derivation of the fluxes associated to
the backgrounds of section 4 
from the generalized metric.
Finally, let us make 
one more short comment on the appearance of non-commutativity and non-associativity in the presence of $b$-field gauge transformations
and
 non-geometric backgrounds.
Performing a world-sheet quantization procedure of the closed string 
coordinates $X^i(\tau,\sigma)$ in non-geometric target spaces, 
one  loses  
commutativity and even associativity of the closed string coordinates (see refs.\cite{Blumenhagen:2010hj,Lust:2010iy,Blumenhagen:2011ph,Condeescu:2012sp,Andriot:2012vb}). It is tempting to speculate that this non-associativity of the closed string geometry, derived from the world-sheet theory, is related to the DFT non-associativity of   
generalized coordinate transformations.
At the moment, however, we do not see a logical connection between these two non-associative structures, but this issue certainly deserves further studies.

We now mention a few other developments and possible generalization 
that we did not cover in more detail in the main text. 
DFT has been applied in various contexts,   
including the heterotic theory \cite{Hohm:2011ex}, massless and massive type II theories \cite{Hohm:2011zr,Hohm:2011cp,Jeon:2012kd}, and  
their supersymmetric extensions \cite{Hohm:2011nu,Coimbra:2011nw,Jeon:2011sq}, and also leads to a compelling 
generalization of Riemannian geometry \cite{Hohm:2010xe,Hohm:2011si,Hohm:2012gk,Hohm:2012mf,Jeon:2010rw}, 
which in turn is closely related to (and an extension of) results in 
 generalized geometry
 \cite{Hitchin:2004ut,Gualtieri:2003dx,Gualtieri:2007bq,Vaisman:2012ke}. 
Another intriguing feature of DFT is that it makes manifest a `double copy' property of supergravity, 
i.e., a certain factorization of gravity Feynman rules in terms of Yang-Mills-like Feynman rules, 
as was already pointed out in  \cite{Siegel:1993th} and shown more explicitly in  \cite{Hohm:2011dz}.  
The ideas of DFT as well as earlier work on 
U-dualities\cite{Hull:2004in,Hull:2007zu} 
have also been extended 
 in order to describe truncations of 
$D=11$ supergravity \cite{Berman:2010is,Berman:2011pe,Thompson:2011uw,Berman:2011cg,Berman:2011jh,Coimbra:2011ky,Berman:2012vc,Cederwall:2013naa,
Cederwall:2013oaa}. 
Until recently, however, they were restricted to rather 
severe truncations of $D=11$ supergravity, but it has now been shown how to formulate complete gravity theories  
in fully U-duality covariant manner in \cite{Hohm:2013jma,Hohm:2013pua}, 
leading to an `exceptional field theory' 
analogue of double field theory. 
Other developments and further results related to 
DFT have appeared in \cite{Kwak:2010ew,Maharana:2010sp,Copland:2011yh,Schulz:2011ye,Dibitetto:2012rk}.  
We finally note that short, early reviews can be found in \cite{Hohm:2011gs}. More extensive 
and more recent
reviews have appeared in~\cite{Newreview}.

\bigskip
We conclude by listing some outstanding questions:

\begin{itemize}

 \item In DFT  the $b$-field gauge transformations are treated geometrically
 as coordinate transformations on the doubled space. 
 The known mathematical framework for $b$-fields   
 is given in the language of `gerbes' and thus it would be 
 useful to understand any relation between these approaches.

 \item How does the compactification on non-geometric backgrounds  
 proceed in general?  It should lead to gauged supergravity 
 theories that so far had no  higher-dimensional ancestor, 
 as was recently discussed in \cite{Condeescu:2013yma} for the non-geometric form given 
 in eq.~(\ref{nongeometaurho}).

 \item There should be a better understanding of the solution space for the
weaker constraints used in the context of the extended DFT that yields
gauged supergravities upon Scherk-Schwarz reduction. 
What kinds of  doubled coordinate dependence do they allow?  
What is the relation to string theory, 
given that the weak constraint, which holds in string theory, is now 
relaxed?
 
 \item Since infinitesimal gauge transformations receive $\alpha'$ deformations,
 the same should happen for finite or large gauge transformations. 
 Can we write a natural, closed-form expression for these corrected large
 gauge transformations?
 
 \item The  $\alpha'$ deformations of the gauge transformations, bracket, and action 
 may be generalized to other string theories. In particular, in the context of superstring theories, it 
 may shed light on the subtle relation 
 between duality symmetries and supersymmetry. 

 \item In the `exceptional field fheory' 
one may investigate the same generalizations and applications 
relevant to DFT --  to supersymmetry, higher derivatives corrections, non-trivial backgrounds, etc.
\end{itemize}

\section*{Acknowledgments}
We would like to thank F.~Hassler, C.~Hull, A.~Kleinschmidt, D.~Marques, H.~Samtleben, W.~Siegel, A.~Sen and 
W.~Taylor for
helpful discussions.

This work is supported by the U.S. Department of Energy (DoE) under the cooperative
research agreement DE-FG02-05ER41360, the
DFG Transregional Collaborative Research Centre TRR 33, 
the DFG cluster of excellence "Origin and Structure of the Universe" and 
by the ERC Advanced Grant �Strings and Gravity� (Grant.No.~32004).

\end{document}